\documentclass[12pt]{article}
\pdfoutput=1

\usepackage[a4paper,text={16.8cm,22.4cm}]{geometry}
\usepackage{amsmath,amsfonts,braket,slashed,amssymb,tikz,bm,psfrag,graphicx,color,dsfont,euscript}
\usepackage{multicol}
\usepackage{ctable}
\usepackage[small,labelfont=bf]{caption}

\RequirePackage[sort&compress,square,comma,numbers]{natbib}
\allowdisplaybreaks
\newcommand{\A}{{\EuScript A}}
\newcommand{\B}{{\EuScript B}}
\newcommand{\X}{{\EuScript X}}
\newcommand{\G}{{\EuScript G}}
\newcommand{\W}{{\EuScript W}}

\newcommand{\nsl}{\rlap{\hspace{0.25mm}/}{n}}
\newcommand{\nbsl}{\rlap{\hspace{0.25mm}/}{\bar n}}
\newcommand{\Asl}{\rlap{\hspace{0.6mm}/}{\A}}
\newcommand{\Dsl}{\rlap{\hspace{0.75mm}/}{D}}
\newcommand{\delsl}{\rlap{\hspace{0.2mm}/}{\partial}}

\newcommand{\arcangle}{\mathord{<\mspace{-9.5mu}\mathrel{)}\mspace{2mu}}}

\begin{document}

\begin{titlepage}

\begin{flushright}
\normalsize
MITP/18-038\\ 
ZH-TH~19/18\\
June 4, 2018
\end{flushright}

\vspace{1.0cm}
\begin{center}
\Large\bf\boldmath
Effective Field Theory after a New-Physics Discovery
\end{center}

\vspace{0.5cm}
\begin{center}
Stefan Alte$^a$, Matthias K\"onig$^b$ and Matthias Neubert$^{a,c}$\\
\vspace{0.7cm} 
{\sl ${}^a$PRISMA Cluster of Excellence \& Mainz Institute for Theoretical Physics\\
Johannes Gutenberg University, 55099 Mainz, Germany\\[3mm]
${}^b$Physik-Institut, Universit\"at Z\"urich, CH-8057, Switzerland\\[3mm]
${}^c$Department of Physics \& LEPP, Cornell University, Ithaca, NY 14853, U.S.A.}
\end{center}

\vspace{0.8cm}
\begin{abstract}
When a new heavy particle is discovered at the LHC or at a future high-energy collider, it will be interesting to study its decays into Standard Model particles using an effective field-theory framework. We point out that the proper effective theory must be based on non-local operators defined in soft-collinear effective theory (SCET). For the interesting case where the new resonance is a gauge-singlet spin-0 boson, which is the first member of a new sector governed by a mass scale $M$, we show how a consistent scale separation between $M$ and the electroweak scale $v$ is achieved up to next-to-next-to-leading order in the expansion parameter $\lambda\sim v/M$. The Wilson coefficients in the effective Lagrangian depend in a non-trivial way on the mass of the new resonance and the masses of yet undiscovered heavy particles. Large logarithms of the ratio $M/v$ can be systematically resummed using the renormalization group. We develop a SCET toolbox, with which it is straightforward to construct the relevant effective Lagrangians for new heavy particles with other charges and spin.
\end{abstract}

\end{titlepage}

\tableofcontents
\newpage

\section{Introduction}

Following the discovery of a new heavy particle with mass far above the electroweak scale, understanding its properties will be a crucial task for both theorists and experimenters. In many well-motivated extensions of the Standard Model (SM), such as models based on supersymmetry, compositeness, or extra dimensions, one expects that the first new particle to be discovered is one member of a larger sector of particles with similar masses, characterized by a scale $M\gg v$. Barring any further discoveries, the most general approach to studying the new particle's properties -- via its decays into SM particles and its production rates -- would be to embed it into an effective field-theory (EFT) formalism. The purpose of this work is to show how this can be done consistently.

While no new particles have yet been discovered at the LHC, the high-luminosity run still offers a significant discovery potential for new heavy resonances, for which the mass reach extends out to about 6\,TeV (see e.g.\ \cite{CMS:2013xfa,ATLAS:2013hta}). An energy upgrade to 27\,TeV or a future 100\,TeV collider could extend this reach significantly. The phantom 750\,GeV diphoton resonance, for which preliminary evidence was reported by the ATLAS and CMS collaborations in late 2015 \cite{Aaboud:2016tru,Khachatryan:2016hje}, provides a concrete example with which to illustrate the motivation for our work. Hundreds of phenomenological papers have been written in response to these hints. In most of them, the authors have assumed the existence of a neutral spin-0 boson $S$ with mass $M_S\approx 750$\,GeV and constructed the most general EFT Lagrangian at dimension-5 order, in which $S$ is coupled to SM fields. The underlying assumption is that these dimension-5 operators arise from integrating out additional heavy particles. However, in the vast majority of models addressing the diphoton resonance these other particles had masses of the same order, governed by a scale $M\sim M_S\gtrsim 1$\,TeV. In such a situation, it is evident that a conventional EFT approach cannot be employed in a systematic way to study the on-shell decay and production rates of the new particle. The naive assumption that amplitudes of the dimension-5 Lagrangian scale like $v^n/M$, where $v\approx 246$\,GeV is the electroweak scale, is invalid in this case. The reason is that EFT matrix elements scale with powers of the mass parameters present in the theory, which now are $v$ and $M_S$. For $M_S\sim M\gg v$, higher-dimensional operators can be unsuppressed with respect to lower-dimensional ones, since their contributions can scale with $(M_S/M)^{2n}={\cal O}(1)$ relative to the dimension-5 contributions. Factors of $M_S^2$ in the numerator can arise, e.g., from operators containing extra derivatives or longitudinally polarized gauge fields. Thus, infinite towers of EFT operators would need to be retained to include all terms of a given order in $v/M_S$ -- a task that is usually impracticable. Also, a conventional EFT would not allow one to resum large logarithms of the scale ratio $M_S/v$.

A successful theoretical framework to address this situation will have to accomplish the following tasks: i) it must be flexible enough to retain the full dependence on the two new-physics scales: the mass $M_S$ of the heavy resonance that has been discovered, and the mass scale $M$ characterizing the other particles belonging to the new sector; ii) it must allow for a consistent separation of the contributions arising from the scales $M_S$ and $v$, and in particular it must provide the tools to resum large (double) logarithms of the scale ratio $M_S/v$ using renormalization-group (RG) equations. Note that with $M_S\sim\mbox{few TeV}$ these logarithms can be very large, e.g.\ $\alpha_s\ln^2(M_S^2/m_t^2)\sim 5$ for $M_S=5$\,TeV, and hence resummation is obligatory, even for electroweak radiative corrections.

The situation encountered here is similar to the case of $B$-meson decays to final states containing light mesons. A systematic heavy-quark expansion of the corresponding decay amplitudes in the small ratio $\Lambda_{\rm QCD}/m_b$ is made complicated by the fact that the light final-state particles carry energies $E_i={\cal O}(m_b)$ that scale with the heavy-quark mass. This obstacle was overcome with the QCD factorization approach developed in \cite{Beneke:1999br,Beneke:2000ry,Beneke:2001ev} and the construction of soft-collinear effective theory (SCET) \cite{Bauer:2000yr,Bauer:2001ct,Bauer:2001yt,Beneke:2002ph}. In the present work, we use established SCET technology to derive a consistent EFT that can be employed to study the decays of a new heavy particle $S$ into SM particles. The decay amplitudes are systematically expanded in powers of the ratio $\lambda=v/M_S\ll 1$. The scale $M_S$ enters via the large energies and momenta carried by the light SM particles in the final state. While SCET was developed for QCD processes originally, generalizations to electroweak processes have been discussed in \cite{Chiu:2007yn,Chiu:2007dg,Chiu:2008vv}. In several aspects our approach follows the line of reasoning laid out in these papers. However, we go significantly further by developing the SCET approach beyond the leading order in the power expansion, where several new and subtle issues arise. For example, there is a non-trivial mixing of operators at leading and subleading order, which gives rise to a novel source of large double logarithms, which we resum. We shall refer to the effective field theory we develop as ``SCET beyond the SM'' (SCET$_{\rm BSM}$). 

We stress that our effective theory is not meant as an alternative to the EFT extension of the SM referred to as SMEFT \cite{Weinberg:1979sa,Wilczek:1979hc,Buchmuller:1985jz,Leung:1984ni,Grzadkowski:2010es} (see \cite{Brivio:2017vri} for a recent review). SMEFT parameterizes new-physics effects from heavy virtual particles in a model-independent way by extending the SM through local, higher-dimensional operators built out of SM fields. Assuming there are no light new particles beyond the SM, it provides the appropriate EFT framework for studying indirect hints of new physics. SCET$_{\rm BSM}$, on the other hand, is constructed to describe the decays of a new on-shell heavy resonance into SM particles. In our treatment we will assume that the new resonance is narrow ($\Gamma_S/M_S\ll 1$), such that its width can be neglected when constructing the effective theory. If $S$ decays primarily into SM particles, our results obtained for the various decay widths show {\em a posteriori\/} that this assumption is justified.

The construction of the SCET$_{\rm BSM}$ Lagrangian is process dependent. In this paper we will develop a general toolbox, which allows for a simple, systematic and intuitive construction of the relevant effective Lagrangians for BSM practitioners, even if they are not experts on SCET. For simplicity, we assume that $S$ has spin-0 and is a gauge singlet under the SM. After reviewing some basic aspects of SCET in Section~\ref{sec:SCETbasics}, we construct in Sections~\ref{sec:SCET} and~\ref{sec:3body} the relevant effective Lagrangians for all two-body decays of $S$ into SM particles, and for all three-body decay processes involving a fermion pair in the final state. The extension to new particles with spin $S=1/2$ or 1, or particles which carry SM quantum numbers, is straightforward. However, if $S$ is a member of an $SU(2)_L$ multiplet, then a gauge-invariant EFT can only be built in terms of the entire multiplet. 

In the conventional EFT approach, the decay amplitudes of $S$ into pairs of SM particles receive contributions from operators of dimension $D=5$ (in the case of $S\to Zh$ these contributions start at one-loop order), but nevertheless these amplitudes have different scaling properties with $\lambda=v/M_S$, namely (see e.g.\ \cite{Franceschini:2015kwy,Bauer:2016zfj}) 
\begin{equation}\label{scalings}
\begin{aligned}
   {\cal M}(S\to hh) &= {\cal O}(\lambda^0) \,, &\qquad
   {\cal M}(S\to VV) &= {\cal O}(\lambda^0) \,, \\
   {\cal M}(S\to f\bar f) &= {\cal O}(\lambda) \,, &
   {\cal M}(S\to Zh) &= {\cal O}(\lambda^2) \,, 
\end{aligned}
\end{equation} 
where $V$ represents a gauge boson (massive or massless) and $f$ a fermion. As mentioned earlier, for $M_S\sim M$ an infinite tower of higher-dimensional operators with $D\ge 7$ can give rise to unsuppressed corrections to these amplitudes. For example, the operators
\begin{equation}\label{ex1}
   \frac{1}{M}\,S B_{\mu\nu} B^{\mu\nu} 
    \quad \mbox{and} \quad
   \frac{1}{M^3}\,S\,(\partial_\alpha B_{\mu\nu}) (\partial^\alpha B^{\mu\nu}) \,,
\end{equation} 
where $B^{\mu\nu}$ denotes the field strength associated with hypercharge, contribute terms of order $M_S^2/M$ and $M_S^4/M^3$ to the $S\to\gamma\gamma$ amplitude, respectively. In the case of the decay $S\to Zh$, the scaling ${\cal M}(S\to Zh)\propto v^2/M$ derived in \cite{Bauer:2016zfj} arose from apparently accidental cancellations of terms scaling like $M_S^2/M$ among different diagrams, and it is thus well motivated to ask whether higher-dimensional operators induce larger contributions scaling like $M_S^{2n}/M^{2n-1}={\cal O}(\lambda^0)$. 

In the present work, we derive the scaling laws (\ref{scalings}) from first principles and show that they remain valid even in the case where the two scales $M$ and $M_S$ are of the same order. To this end, we construct the relevant SCET$_{\rm BSM}$ Lagrangians up to next-to-next-to-leading order (NNLO) in $\lambda$. The finite sets of non-local SCET operators arising at each order in the $\lambda$ expansion accounts for infinite towers of local EFT operators. The scaling properties of the operators in SCET translate directly into the scalings of the various decay amplitudes. The complete information about the ultra-violet (UV) completion of the theory, i.e.\ about the yet unknown particles with masses of order $M\sim M_S$ and their interactions, is contained in the Wilson coefficients of the effective Lagrangian. In Section~\ref{sec:RGEs} we show how by solving RG equations one can resum the large (double) logarithms of the scale ratio $M_S/v$. While most of our discussion focusses on the interesting case where $M\sim M_S$ are two scales of the same order, we discuss in Section~\ref{sec:hierarchy} scenarios in which there is a double hierarchy, such that $M\gg M_S\gg v$. In this case a conventional EFT framework can be used to identify the leading terms in an expansion in powers of $M_S/M$, while the SCET$_{\rm BSM}$ is needed to organize in a systematic way the expansion in $\lambda=v/M_S$ and resum large logarithms of this scale ratio. We derive model-independent expressions for the Wilson coefficients in the SCET$_{\rm BSM}$ Lagrangian in terms of the parameters of the local EFT including operators up to dimension~5. In Section~\ref{sec:concl} we present our conclusions along with an outlook on future work.

\section{Basic elements of SCET}
\label{sec:SCETbasics}

Our goal in this work is to develop a consistent EFT for the analysis of the on-shell decays of a hypothetical new, heavy spin-0 boson $S$ (with mass $M_S\gg v$) into SM particles. For simplicity we assume that $S$ is a singlet under the SM gauge group. We also allow for the existence of other heavy particles with similar masses $M\sim M_S$, which have not yet been discovered. They are integrated out and thus do not appear as degrees of freedom in the effective Lagrangian. As we will show, the appropriate EFT is intrinsically non-local and consists of operators defined in SCET. Nevertheless, the theory is well defined and can be constructed following a set of simple rules. As our desire is to elucidate the main ideas of our proposal and to present the construction of the SCET$_{\rm BSM}$ Lagrangian in the most simple and transparent way, we will be brief on some technicalities, which are familiar to SCET practitioners but may look intimidating to others. Interested readers can find more details in the original papers \cite{Bauer:2000yr,Bauer:2001ct,Bauer:2001yt,Beneke:2002ph} and in the review \cite{Becher:2014oda}.

The intrinsic complication in constructing an EFT for the decays of a heavy particle $S$ into light (or massless) particles is that the large mass $M_S$ enters the low-energy theory as a parameter characterizing the large energies $E_i\sim M_S$ of the final-state particles. This is different from conventional EFTs of the Wilsonian type, in which short-distance fluctuations of heavy virtual particles are integrated out from the generating functional of low-energy Green's functions. In SCET, the large energies carried by the light particles give rise to non-localities along the nearly light-like directions in which these particles travel. 

In a given decay process of the heavy particle $S$, the final state contains jets defining directions $\{\bm{n}_1,\dots,\bm{n}_k\}$ of large energy flow. Each jet may consist of one or more collinear particles, which have energies much larger than their rest masses. For each jet direction $\bm{n}_i$, we define two light-like reference vectors $n_i^\mu=(1,\bm{n}_i)$ and $\bar n_i^\mu=(1,-\bm{n}_i)$, with $n_i\cdot\bar n_i=2$. The 4-momentum $p$ of a particle in the jet can then be written as
\begin{equation}\label{pexp}
   p^\mu = \bar n_i\cdot p\,\frac{n_i^\mu}{2} + n_i\cdot p\,\frac{\bar n_i^\mu}{2} + p_\perp^\mu \,,
\end{equation}
where $\bar n_i\cdot p={\cal O}(M_S)$ is much larger than $n_i\cdot p={\cal O}(m^2/M_S)$. The different components scale as 
\begin{equation}
\label{scaling}
   (n_i\cdot p,\bar n_i\cdot p,p_\perp) \sim M_S\,(\lambda^2,1,\lambda) \,,
\end{equation}
where $\lambda=v/M_S$ is the expansion parameter of the effective theory, and we assume that the masses of the light particles are set by the electroweak scale $v$. Particles whose momenta scale in this way are referred to as ``$n_i$-collinear particles''. The particles inside a given jet can interact with each other according to the Feynman rules of SCET, which are equivalent to the usual Feynman rules of the SM \cite{Beneke:2002ph}. However, an $n_i$-collinear particle cannot interact directly with an $n_j$-collinear particle contained in another jet.\footnote{Such interactions can however be mediated by the exchange of ultra-soft particles, see Section~\ref{sec:RGEs}.} 
The effective Lagrangian of SCET, from which one derives the Feynman rules, is discussed in the Appendix.

In SCET, $n_i$-collinear particles are described by effective fields referred to as ``collinear building blocks'' \cite{Bauer:2002nz,Hill:2002vw}. They are composite fields invariant under so-called ``$n_i$-collinear gauge transformations'', which preserve the scaling of the particle momenta shown in (\ref{scaling}). The building blocks are defined with the help of $n_i$-collinear Wilson lines \cite{Bauer:2000yr,Bauer:2001ct,Bauer:2001yt} built out of the various gauge bosons associated with the SM gauge group. We define 
\begin{equation}\label{Wn}
\begin{aligned}
   W_{n_i}^{(G)}(x) &= P\exp\left[ ig_s \int_{-\infty}^0\!ds\,
    \bar n_i\cdot G_{n_i}(x+s\bar n_i) \right] , \\
   W_{n_i}^{(W)}(x) &= P\exp\left[ ig \int_{-\infty}^0\!ds\,
    \bar n_i\cdot W_{n_i}(x+s\bar n_i) \right] , \\
   W_{n_i}^{(B)}(x) &= P\exp\left[ ig'\,Y\! \int_{-\infty}^0\!ds\,
    \bar n_i\cdot B_{n_i}(x+s\bar n_i) \right] , \\
\end{aligned}
\end{equation}
where $g_s$, $g$ and $g'$ denote the gauge couplings of $SU(3)_c$, $SU(2)_L$ and $U(1)_Y$, while $G_{n_i}^\mu(x)\equiv G_{n_i}^{\mu,a}(x)\,t^a$, $W_{n_i}^\mu(x)\equiv W_{n_i}^{\mu,a}(x)\,\tau^a$ and $B_{n_i}(x)$ denote the corresponding $n_i$-collinear gauge fields. They are defined such that their Fourier transforms only contain particle modes whose momenta satisfy the scaling in (\ref{scaling}). The path-ordering symbol ``$P$'' is defined such that the gauge fields are ordered from left to right in order of decreasing $s$ values. For a given SM field, the corresponding collinear Wilson line is obtained by the appropriate product of the objects defined in (\ref{Wn}), where the hypercharge generator $Y$ in the definition of $W_{n_i}^{(B)}$ is replaced by the hypercharge of the respective field. For example, the collinear Wilson lines for the scalar Higgs doublet and a right-handed up-quark field are
\begin{equation}
   W_{n_i}(x) = W_{n_i}^{(W)}(x)\,W_{n_i}^{(B)}(x) \quad \mbox{and} \quad
   W_{n_i}(x) = W_{n_i}^{(G)}(x)\,W_{n_i}^{(B)}(x) \,,
\end{equation}
where $Y$ takes the values $\frac12$ and $\frac23$, respectively. 

The $n_i$-collinear building blocks for the scalar Higgs doublet and the SM fermions are defined as \begin{equation}\label{calXdef}
\begin{aligned}
   \Phi_{n_i}(x) &= W_{n_i}^\dagger(x)\,\phi(x) \,, \\
   \X_{n_i}(x) &= \frac{\nsl_i\nbsl_i}{4}\,W_{n_i}^\dagger(x)\,\psi(x)
    \equiv P_{n_i} W_{n_i}^\dagger(x)\,\psi(x) \,,
\end{aligned}
\end{equation}
where the projection operator $P_{n_i}$, which is defined such that $\nsl_i P_{n_i}=0$ and $P_{n_i}^2=P_{n_i}$, projects out the large components of the spinor of a highly energetic fermion. The $n_i$-collinear building blocks for the gauge bosons are defined as (for $A=G,W,B$) \cite{Bauer:2002nz,Hill:2002vw}
\begin{equation}\label{calAdef}
   \A_{n_i}^\mu(x) = W_{n_i}^{(A)\dagger}(x)\,\big[ iD_{n_i}^\mu\,W_{n_i}^{(A)}(x) \big] 
   = g_A \int_{-\infty}^0\!ds\,{\bar n}_i{}_\nu \big[ 
    W_{n_i}^{(A)\dagger} F_{n_i}^{\nu\mu}\,W_{n_i}^{(A)} \big](x+s\bar n_i) \,, 
\end{equation}
where $iD_{n_i}^\mu=i\partial^\mu+g_A A_{n_i}^\mu$ denotes the collinear covariant derivative, $g_A$ is the appropriate gauge coupling (which in the case $A=B$ includes the hypercharge generator, so $g_G\equiv g_s$, $g_W\equiv g$, and $g_B\equiv g'\,Y$), and $F_{n_i}^{\nu\mu}$ is the field-strength tensor associated with the collinear gauge field $A_{n_i}^\mu$. Note that for the hypercharge gauge field the Wilson lines cancel out in the last expression in (\ref{calAdef}), and hence one finds 
\begin{equation}
   \B_{n_i}^\mu(x) 
   = g'\,Y \int_{-\infty}^0\!ds\,{\bar n}_i{}_\alpha B_{n_i}^{\alpha\mu}(x+s\bar n_i) \,.
\end{equation}
We will also use the expansions of the gauge-boson building blocks in the generators of the gauge groups, i.e.\  
\begin{equation}
   \G_{n_i}^\mu(x) = \G_{n_i}^{\mu,a}(x)\,t^a \,, \qquad
   \W_{n_i}^\mu(x) = \W_{n_i}^{\mu,a}(x)\,\tau^a \,, \qquad
   \B_{n_i}^\mu(x) = Y\,\B_{n_i}^{\mu,a}(x) \,, 
\end{equation}
where in the latter case $a=1$. The building blocks for the collinear fermion and gauge fields satisfy the constraints
\begin{equation}
   \nsl_i\,\X_{n_i}(x) = 0 \,, \qquad \bar n_i\cdot\A_{n_i}(x) = 0 \,.
\end{equation}
The Wilson lines contain the longitudinal components $\bar n_i\cdot A_{n_i}$ of the gauge fields, while the gauge-invariant collinear fields $\A_{n_i}^\mu$ themselves have no such components. Because of the presence of the Wilson lines, the SCET fields can create or absorb particles along with an arbitrary number of (longitudinal) gauge bosons coupling to these particles and traveling in the same direction. In this sense the effective fields describe ``jets'' of collinear partons. Note that a different set of collinear fields (scalars, fermions and gauge fields) is introduced for each direction $\bm{n}_i$ of large energy flow. 

The collinear building blocks have well-defined scaling properties with the expansion parameter $\lambda$. One finds \cite{Bauer:2001yt,Beneke:2002ph}
\begin{equation}\label{counting}
   \Phi_{n_i} \sim \lambda \,, \qquad
   \X_{n_i} \sim \lambda \,, \qquad
   \A_{n_i\perp}^\mu \sim \lambda \,, \qquad
   n_i\cdot\A_{n_i} \sim \lambda^2 \,.
\end{equation}
In analogy with (\ref{pexp}), the transverse gauge fields are defined as
\begin{equation}\label{Aperpdef}
   \A_{n_i\perp}^\mu = \A_{n_i}^\mu - n_i\cdot\A_{n_i}\,\frac{\bar n_i^\mu}{2} \,,
\end{equation}
where we have used that $\bar n_i\cdot\A_{n_i}=0$.

It follows that operators containing $N$ collinear fields (irrespective of their directions) have scaling dimension $d\ge N$ in $\lambda$, and adding more fields to an operator always increases its scaling dimension. This is how SCET can be employed to construct a consistent expansion in powers of $\lambda$. Operators in the effective Lagrangian can also contain derivatives acting on collinear fields, which produce collinear momenta when taking matrix elements of an operator. There is no need to use covariant derivatives, since the building blocks are gauge invariant by themselves. From (\ref{scaling}) it follows that one can add an arbitrary number of $i\bar n_i\cdot\partial$ derivatives acting on $n_i$-collinear fields, while $i n_i\cdot\partial$ or $i\partial_\perp^\mu$ derivatives gives rise to additional power suppression. The freedom to introduce $i\bar n_i\cdot\partial$ derivatives at will implies that $n_i$-collinear fields can be delocalized along the $\bar n_i$ direction, and hence the operators appearing in the SCET Lagrangian are non-local. A first hint at this non-locality is the presence of the Wilson lines themselves, see (\ref{Wn}).

The heavy particle $S$ should be represented in the effective theory by an effective field $S_v(x)\,e^{-iM_S v\cdot x}$, whose soft interactions are described by a ``heavy-particle effective theory'' constructed in analogy with heavy-quark effective theory \cite{Eichten:1989zv,Georgi:1990um,Eichten:1990vp,Falk:1990yz,Falk:1990pz,Neubert:1993mb}. Since in our case $S$ is a gauge singlet and has no interactions, this step is unnecessary. It would become a relevant step if one constructs the effective theory for a resonance $S$ that is charged under any of the SM gauge groups.

\section{\boldmath SCET$_{\rm BSM}$ for two-body decays of $S$}
\label{sec:SCET}

\begin{figure}
\begin{center}
\includegraphics[width=0.3\textwidth]{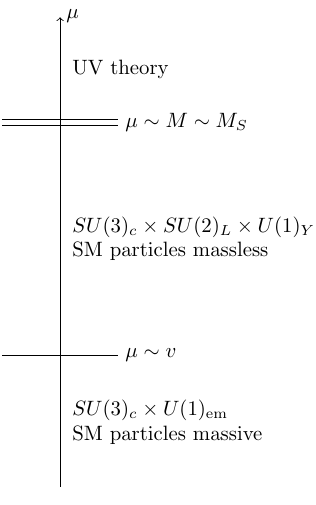} \qquad\qquad
\includegraphics[width=0.1885\textwidth]{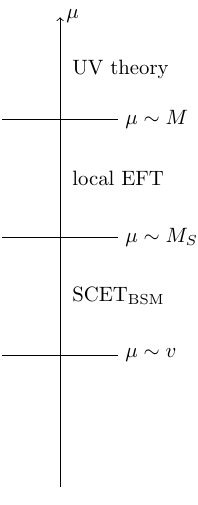}
\end{center}
\vspace{-3mm}
\caption{\label{fig:scales} 
Schematic description of the construction of the SCET$_{\rm BSM}$ for the generic case $M\sim M_S$ (left), and for the case of a double hierarchy $M\gg M_S\gg v$ (right).}
\end{figure}

We now have the tools to construct an EFT for the decays of a new heavy particle $S$ with mass $M_S\gg v$ into SM particles. The basic construction of the SCET$_{\rm BSM}$ is illustrated in the left panel of Figure~\ref{fig:scales}. It consists of the following steps:
\begin{enumerate}
\item 
At the new-physics scale $\mu\sim M_S\sim M$, the complete UV theory (which is unknown, of course) is matched onto an extension of SCET built out of the resonance $S$ and $n_i$-collinear SM fields. All heavy particles besides the resonance $S$, as well as ``hard'' quantum fluctuations with virtualities of order $M_S$, are integrated out in this step. Since the mass of $S$ is much above the electroweak scale, its interactions can be described in terms of operators in the unbroken phase of the electroweak symmetry, preserving full $SU(3)_c\times SU(2)_L\times U(1)_Y$ gauge invariance. If there is a hierarchy between the scales $M$ and $M_S$ (right panel of Figure~\ref{fig:scales}), then the two scales are integrated out in two steps, see Section~\ref{sec:hierarchy}. 
\item 
In the next step, the effective operators and their Wilson coefficients are evolved from the high-energy scale $\mu\sim M_S$ to the electroweak scale $\mu\sim v$. This is accomplished by solving the RG equations of the effective theory. In this process, the SM particles can be treated as massless. In the SCET community, this version of the effective theory is called SCET$_{\rm I}$. The relevant anomalous dimensions can be calculated using standard technology. Solving the RGEs resums large logarithms of the scale ratio $M_S/v$ to all orders in perturbation theory.\footnote{Unlike in applications of SCET to hadronic decays of $B$ mesons, there is no need to perform an additional matching at an intermediate ``hard-collinear'' scale $\mu\sim\sqrt{v M_S}$ \cite{Chiu:2008vv}. The reason is simply that no such scale can be formed out of the physical momenta of the particles involved in the decay.}
\item
At the electroweak scale the symmetry is broken to $SU(3)_c\times U(1)_{\rm em}$, and mass effects from SM particles need to be taken into account. This is accomplished by introducing mass terms for the $n_i$-collinear fields. In loop calculations, it is also necessary to include so-called soft mass-mode fields with momentum scaling $(\lambda,\lambda,\lambda)$ \cite{Fleming:2007qr,Fleming:2007xt,Chiu:2009yx}. This version of the effective theory is often referred to as SCET$_{\rm II}$. The presence of mass terms in loop calculations gives rise to the collinear anomaly \cite{Becher:2010tm}. The corresponding loop integrals require an additional analytic regulator beyond dimensional regularization, which leads to the appearance of additional large logarithms in the matrix elements of the low-energy effective theory. It can be shown that these rapidity logarithms do not exponentiate and hence they do not spoil the resummation accomplished in step~2 \cite{Chiu:2007dg,Becher:2010tm,Chiu:2011qc}.
\item
If one is interested in processes involving particles much lighter than the weak scale, then at $\mu\approx v$ an additional matching step is required, in which the SM particles with weak-scale masses (the top quark, the Higgs boson, and the $W$ and $Z$ bosons) are integrated out. This theory is then evolved down to a scale $\mu$ characteristic to the process of interest, where the relevant operator matrix elements are evaluated. 
\end{enumerate}

Each $n_i$-collinear field in the SCET$_{\rm BSM}$ Lagrangian carries a collinear momentum in the corresponding direction $\bm{n}_i$ with a large net energy and thus must produce at least one $n_i$-collinear particle entering the final state. By momentum conservation, each operator in the SCET$_{\rm BSM}$ Lagrangian must contain at least two different types of collinear fields, representing particles moving in different directions. Because of electroweak symmetry breaking, the effective theory also contains scalar fields carrying no 4-momentum. These are represented by a constant field $\Phi_0\sim\lambda$, which does not transform under collinear gauge transformations. After electroweak symmetry breaking one replaces
\begin{equation}
   \Phi_0 \,\stackrel{\rm EWSB}{\to}\, 
   \frac{1}{\sqrt2} \left( \begin{array}{c} 0 \\ v \end{array} \right) .
\end{equation}

In this section we focus on the simplest, but phenomenologically most important case of two-body decays of the heavy resonance $S$. Then the vectors $\bm{n}_2=-\bm{n}_1$ point in opposite directions, and therefore $n_2=\bar n_1$ and $n_1=\bar n_2$ for the light-like reference vectors. Since the choice of the direction of the reference vectors is arbitrary, all operators in the effective Lagrangian must be invariant under the exchange $n_1\leftrightarrow n_2$. 

\subsection[Effective Lagrangian at ${\cal O}(\lambda^2)$]{\boldmath Effective Lagrangian at ${\cal O}(\lambda^2)$}
\label{sec:3.1}

It is convenient to work in the rest frame of the decaying particle, in which the light final-state particles carry large energies $E_i={\cal O}(M_S)$. Since the operators in the effective Lagrangian must contain at least one $n_1$-collinear and one $n_2$-collinear field, the power-counting rules in (\ref{counting}) imply that the leading operators have scaling dimension $d=2$. While invariance under $n_i$-collinear gauge transformations is ensured by constructing the effective Lagrangian in terms of collinear building blocks, the operators must also be invariant under global gauge transformations, i.e.\ they must conserve the color and electroweak charges. At ${\cal O}(\lambda^2)$, the only gauge-invariant operators are those containing either two scalar doublets or two transverse gauge fields. Considering the first possibility, we write the corresponding term in the effective Lagrangian as
\begin{equation}
   {\cal L}_{\rm eff} \ni \int\!ds\,dt\,\bar C_{\phi\phi}(s,t,M,\mu)\,
   S(x) \big[ \Phi_{n_1}^\dagger(x+s\bar n_1)\,\Phi_{n_2}(x+t\bar n_2)
    + \Phi_{n_2}^\dagger(x+t\bar n_2)\,\Phi_{n_1}(x+s\bar n_1) \big] \,,
\end{equation} 
where we have taken into account that collinear field operators can be delocalized along the $\bar n_i$ directions, as discussed in Section~\ref{sec:SCETbasics}. The position-space Wilson coefficient $\bar C_{\phi\phi}$ depends on the new-physics scale $M$ via the masses of the yet unknown particles, which have been integrated out, and on the scale $\mu$ at which the effective operator is renormalized. It also depends on the coordinates $s$ and $t$ parameterizing the non-locality of the operator with respect to the position of the field $S(x)$.

The large components $\bar n_i\cdot P_i$ of the total collinear momenta in each jet are fixed by external kinematics. We introduce momentum operators $\bar n_i\cdot{\cal P}_i$ to obtain these components from the quantum fields.\footnote{In some formulations of SCET the collinear fields carry the large momentum components as labels, and the operators $\bar n_i\cdot{\cal P}_i$ are referred to as the ``label operators'' \cite{Bauer:2001yt}.} 
We can then use translational invariance to make the dependence on these components explicit. This gives
\begin{equation}\label{example}
   {\cal L}_{\rm eff} \ni C_{\phi\phi}(\bar n_1\cdot{\cal P}_1,\bar n_2\cdot{\cal P}_2,M,\mu)\,
    S(x)\,\big[ \Phi_{n_1}^\dagger(x)\,\Phi_{n_2}(x)
    + \Phi_{n_2}^\dagger(x)\,\Phi_{n_1}(x) \big] \,,
\end{equation} 
where the Fourier-transformed Wilson coefficient is defined as
\begin{equation}\label{FTC}
   C_{\phi\phi}(\omega_1,\omega_2,M,\mu) 
   = \int\!ds\,dt\,\bar C_{\phi\phi}(s,t,M,\mu)\,e^{is\omega_1}\,e^{it\omega_2} \,.
\end{equation} 
The dependence of the Wilson coefficient on its arguments is restricted by the fact that the Lagrangian must be invariant under the reparameterization transformations $n_i^\mu\to\alpha_i\,n_i^\mu$, $\bar n_i^\mu\to\bar n_i^\mu/\alpha_i$ applied to the light-like reference vectors in each collinear sector \cite{Manohar:2002fd}. It follows that $C_{\phi\phi}$ in (\ref{example}) depends on its first two arguments only through the combination
\begin{equation}\label{eq16}
   \frac{n_1\cdot n_2}{2}\,\bar n_1\cdot{\cal P}_1\,\bar n_2\cdot{\cal P}_2
   = \left( \frac{n_1}{2}\,\bar n_1\cdot{\cal P}_1 + \frac{n_2}{2}\,\bar n_2\cdot{\cal P}_2 \right)^2
   \simeq {\cal P}_S^2 \,. 
\end{equation}
Here and below we use the symbol ``$\simeq$'' for equations valid at leading power in $\lambda$. For two-body decays, the total collinear momenta add up to the momentum of the decaying resonance $S$, and the operator ${\cal P}_S^2$ has eigenvalue $M_S^2$. With a slight abuse of notation, we thus write the corresponding contribution to the effective Lagrangian in the form
\begin{equation}
   {\cal L}_{\rm eff}\ni M\,C_{\phi\phi}(M_S,M,\mu)\,O_{\phi\phi}(\mu) \,, 
    \quad \mbox{with} \quad
   O_{\phi\phi} = S\,\big( \Phi_{n_1}^\dagger \Phi_{n_2} + \Phi_{n_2}^\dagger \Phi_{n_1} \big) \,.
\end{equation} 
All fields are now evaluated at the same spacetime point. We have factored out the new-physics scale $M$ in the final definition of the Wilson coefficient to ensure that the function $C_{\phi\phi}(M_S,M,\mu)$ is dimensionless. Contrary to a conventional EFT, in our approach the short-distance Wilson coefficients depend on all the relevant heavy scales in the problem ($M_S$ and the mass scale $M$ of yet undiscovered heavy particles), and this dependence can be arbitrarily complicated depending on the details of the underlying UV theory. In this way, the SCET$_{\rm BSM}$ Lagrangian accounts for infinite towers of local operators in the conventional EFT approach.

The remaining operators arising at ${\cal O}(\lambda^2)$ contain two transverse gauge fields. Their Lorentz indices can be contracted with the help of two rank-2 tensors defined in the plane transverse to the vectors $n_1$ and $n_2$. We introduce the objects (with $\epsilon_{0123}=-1$)
\begin{equation}\label{gepsdef}
   g_{\mu\nu}^\perp 
   = g_{\mu\nu} - \frac{n_1{}_\mu n_2{}_\nu + n_2{}_\mu n_1{}_\nu}{n_1\cdot n_2} \,, \qquad
   \epsilon_{\mu\nu}^\perp 
   = \epsilon_{\mu\nu\alpha\beta}\,\frac{n_1^\alpha\,n_2^\beta}{n_1\cdot n_2} \,. 
\end{equation}
The latter definition is such that $\epsilon_{12}^\perp=1$ if $n_1^\mu=(1,0,0,1)$ and $n_2^\mu=(1,0,0,-1)$. The complete effective Lagrangian can then be written in the form 
\begin{equation}\label{Leff2}
   {\cal L}_{\rm eff}^{(2)}
   = M\,C_{\phi\phi}(M_S,M,\mu)\,O_{\phi\phi}(\mu) + M\!\!\!\sum_{A=G,W,B}\!\Big[
    C_{AA}(M_S,M,\mu)\,O_{AA}(\mu) + \widetilde C_{AA}(M_S,M,\mu)\,\widetilde O_{AA}(\mu) \Big] ,
\end{equation} 
where (a summation over the group index $a$ is understood for non-abelian fields)
\begin{equation}\label{lam2ops}
\begin{aligned}
   O_{\phi\phi} &= S\,\big( \Phi_{n_1}^\dagger \Phi_{n_2} + \Phi_{n_2}^\dagger \Phi_{n_1} \big) \,, \\
   O_{AA} &= S\,g_{\mu\nu}^\perp\,\A_{n_1}^{\mu,a}\,\A_{n_2}^{\nu,a} \,, \\
   \widetilde O_{AA} &= S\,\epsilon_{\mu\nu}^\perp\,\A_{n_1}^{\mu,a}\,\A_{n_2}^{\nu,a} \,.
\end{aligned}
\end{equation} 
Note that $\epsilon_{\mu\nu}^\perp$ changes sign under $n_1\leftrightarrow n_2$, and hence the last operator indeed has the correct symmetry properties. The first two operators in this list are even under a CP transformation whereas the third operator is odd (assuming that $S$ does not transform under CP). Here and below we indicate CP-odd operators and their Wilson coefficients by a tilde.

The gauge fields contained in the Wilson lines entering the definitions of the gauge-invariant building blocks in (\ref{calXdef}) and (\ref{calAdef}) become important in loop calculations or in applications with multiple emissions of particles in the same jet direction. An exception is the Wilson line associated with the scalar doublet in (\ref{calXdef}), which after electroweak symmetry breaking accounts for the longitudinal polarization states of the physical $W^\pm$ and $Z^0$ bosons.

The SCET$_{\rm BSM}$ Lagrangian (\ref{Leff2}), which is valid for scales $\mu<M_S$, is constructed in the unbroken phase of the electroweak gauge symmetry, in which all particles other than the heavy resonance $S$ can be treated as massless. As shown in Figure~\ref{fig:scales}, at the electroweak scale $\mu\sim v$ this Lagrangian must be matched onto an effective Lagrangian constructed in the broken phase, where the residual gauge symmetry is $SU(3)_c\times U(1)_{\rm em}$ and where the SM particles acquire masses. While this matching is non-trivial at loop order (see e.g.\ \cite{Fleming:2007qr,Fleming:2007xt,Chiu:2007dg,Chiu:2008vv,Chiu:2009yx}), at tree level one simply needs to transform the various fields to the mass basis. In particular, after electroweak symmetry breaking the collinear building block representing the scalar doublet takes the form 
\begin{equation}
   \Phi_{n_i}(0) = \frac{1}{\sqrt2}\,W_{n_i}^\dagger(0)
    \left( \begin{array}{c} 0 \\ v + h_{n_i}(0) \end{array} \right) ,
\end{equation} 
where
\begin{equation}
   W_{n_i}(0) = P\exp\left[ \frac{ig}{2} \int_{-\infty}^0\!ds
    \left( \begin{array}{ccc} \frac{c_w^2-s_w^2}{c_w}\,\bar n_i\cdot Z_{n_i} 
     \!+ 2s_w\,\bar n_i\cdot A_{n_i} &~& \sqrt{2}\,\bar n_i\cdot W_{n_i}^+ \\
     \sqrt{2}\,\bar n_i\cdot W_{n_i}^- && - \frac{1}{c_w}\,\bar n_i\cdot Z_{n_i} \\ 
     \end{array} \right)(s\bar n_i) \right] .
\end{equation}
We have replaced the gauge fields $W^{\mu,a}$ and $B^\mu$ in terms of the mass eigenstates $W^\pm$, $Z$ and $A$. Here $c_w=\cos\theta_W$ and $s_w=\sin\theta_W$ denote the cosine and sine of the weak mixing angle. It follows that
\begin{equation}\label{Opp}
\begin{aligned}
   O_{\phi\phi} &= S(0)\,h_{n_1}(0)\,h_{n_2}(0) 
    + m_Z^2 \int_{-\infty}^0\!ds \int_{-\infty}^0\!dt\,
    S(0)\,\bar n_1\cdot Z_{n_1}(s\bar n_1)\,\bar n_2\cdot Z_{n_2}(t\bar n_2) \\
   &\quad\mbox{}+ m_W^2 \int_{-\infty}^0\!ds \int_{-\infty}^0\!dt\,S(0)
    \left[ \bar n_1\cdot W_{n_1}^-(s\bar n_1)\,\bar n_2\cdot W_{n_2}^+(t\bar n_2)
    + (+\leftrightarrow -) \right] + \dots \,,
\end{aligned}
\end{equation} 
where the dots represent terms containing more than two collinear fields. Taking into account that external collinear Higgs and vector bosons have power counting $\lambda^{-1}$, it follows from (\ref{Leff2}) that the $S\to hh$ and $S\to VV$ decay amplitudes obey the scaling rules shown in  (\ref{scalings}). Note, however, that whereas these rules were obtained by considering dimension-5 operators in the conventional EFT Lagrangian, the scaling relations derived in SCET are exact.

It is straightforward to evaluate the relevant two-body decay amplitudes and decay rates described by the effective Lagrangian (\ref{Leff2}). For the di-Higgs decay mode of $S$, we obtain
\begin{equation}
   {\cal M}(S\to hh) = M\,C_{\phi\phi} \,, \qquad
   \Gamma(S\to hh) = \frac{M^2}{32\pi M_S}\,|C_{\phi\phi}|^2\,\sqrt{1-\frac{4m_h^2}{M_S^2}} \,,
\end{equation}
where here and below we suppress the arguments of the Wilson coefficients. 

The decay amplitudes involving two vector bosons in the final state can be expressed in terms of the general form-factor decomposition 
\begin{equation}\label{VVampl}
   {\cal M}(S\to V_1 V_2) 
   = M \Big[ F_\perp^{V_1 V_2}\,\varepsilon_{1\perp}^*\cdot\varepsilon_{2\perp}^* 
    + \widetilde F_\perp^{V_1 V_2}\,\epsilon_{\mu\nu}^\perp\,\varepsilon_{1\perp}^{*\mu}\,\varepsilon_{2\perp}^{*\nu} 
    + F_\parallel^{V_1 V_2}\,\frac{m_1 m_2}{k_1\cdot k_2}\,
    \varepsilon_{1\parallel}^*\cdot\varepsilon_{2\parallel}^* \Big] \,,
\end{equation}
where $k_i^\mu$ are the momenta of the outgoing bosons, $m_i$ denote their masses, and $\varepsilon_i^\mu\equiv\varepsilon^\mu(k_i)$ are their polarization vectors. The transverse and longitudinal projections of the polarization vectors are defined as
\begin{equation}
   \varepsilon_\perp^\mu(k_i) 
   = \varepsilon^\mu(k_i) - \bar n_i\cdot\varepsilon(k_i)\,\frac{n_i^\mu}{2}
    - n_i\cdot\varepsilon(k_i)\,\frac{\bar n_i^\mu}{2} \,, \qquad 
   \varepsilon_\parallel^\mu(k_i) = \varepsilon^\mu(k_i) - \varepsilon_\perp^\mu(k_i) \,.
\end{equation}
The first two terms in (\ref{VVampl}) correspond to the perpendicular polarization states of the two bosons, while the third term refers to the longitudinal polarization states. The latter only arise for the massive vector bosons $Z^0$ and $W^\pm$. The ratio $m_1 m_2/(k_1\!\cdot\! k_2)$ factored out in the definition of the longitudinal form factor $F_\parallel^{VV}$ takes into account that the longitudinal polarization vectors scale as $\varepsilon_{i\parallel}^\mu(k_i)\simeq k_i^\mu/m_i={\cal O}(\lambda^{-1})$. Our definition ensures that all three form factors are of the same order in SCET power counting. The result (\ref{VVampl}) can also be written in the equivalent form 
\begin{equation}
\begin{aligned}
   {\cal M}(S\to V_1 V_2) 
   &= M F_\perp^{V_1 V_2} \left( \varepsilon_1^*\cdot\varepsilon_2^* 
     - \frac{k_2\cdot\varepsilon_1^*\,k_1\cdot\varepsilon_2^*}%
            {k_1\cdot k_2-\frac{m_1^2\,m_2^2}{k_1\cdot k_2}} \right)
    + M \widetilde F_\perp^{V_1 V_2}\,
    \frac{\epsilon_{\mu\nu\alpha\beta}\,k_1^\mu\,k_2^\nu\, 
          \varepsilon_1^{*\alpha}\,\varepsilon_2^{*\beta}}%
         {\big[(k_1\cdot k_2)^2-m_1^2\,m_2^2\big]^{1/2}} \\
   &\quad\mbox{}+ M F_\parallel^{V_1 V_2}\,
    \frac{m_1 m_2\,k_2\cdot\varepsilon_1^*\,k_1\cdot\varepsilon_2^*}%
         {(k_1\cdot k_2)^2-m_1^2\,m_2^2} \,,
\end{aligned}
\end{equation}
which is independent of the light-like reference vectors used in SCET.

To derive the tree-level expressions for the form factors from the effective Lagrangian (\ref{Leff2}), we use that the one-boson Feynman rule for the gauge-invariant SCET field $\A_{n_i\perp}^{\mu,a}$ yields $g_A\,\varepsilon_{i\perp}^{*\mu}(k_i)$, where $g_A$ denotes the appropriate gauge coupling, while the Wilson-line terms in (\ref{Opp}) produce the structure 
\begin{equation}
   \frac{\bar n_1\cdot\varepsilon_1^*}{\bar n_1\cdot k_1}\,
    \frac{\bar n_2\cdot\varepsilon_2^*}{\bar n_2\cdot k_2}
   = \frac{\varepsilon_{1\parallel}^*\cdot\varepsilon_{2\parallel}^*}{k_1\cdot k_2} \,.
\end{equation}
We thus obtain the transverse form factors
\begin{equation}
\begin{aligned}
   F_\perp^{gg} &= g_s^2\,C_{GG} \,, 
   &\widetilde F_\perp^{gg} &= g_s^2\,\widetilde C_{GG} \,, \\
   F_\perp^{\gamma\gamma} &= e^2 \left( C_{WW} + C_{BB} \right) , 
   &\widetilde F_\perp^{\gamma\gamma} &= e^2 \big( \widetilde C_{WW}
    + \widetilde C_{BB} \big) \,, \\
   F_\perp^{\gamma Z} &= e^2 \left( \frac{c_w}{s_w}\,C_{WW} - \frac{s_w}{c_w}\,C_{BB} \right) , 
   &\widetilde F_\perp^{\gamma Z} 
   &= e^2 \left( \frac{c_w}{s_w}\,\widetilde C_{WW}
    - \frac{s_w}{c_w}\,\widetilde C_{BB} \right) , \\
   F_\perp^{ZZ} &= e^2 \left( \frac{c_w^2}{s_w^2}\,C_{WW} + \frac{s_w^2}{c_w^2}\,C_{BB} \right) ,  
    \qquad
   &\widetilde F_\perp^{ZZ} 
   &= e^2 \left( \frac{c_w^2}{s_w^2}\,\widetilde C_{WW}
    + \frac{s_w^2}{c_w^2}\,\widetilde C_{BB} \right) , \\
   F_\perp^{WW} &= \frac{e^2}{s_w^2}\,C_{WW} , 
   &\widetilde F_\perp^{WW} &= \frac{e^2}{s_w^2}\,\widetilde C_{WW} \,,    
\end{aligned}
\end{equation}
while the longitudinal form factors are given by 
\begin{equation}
   F_\parallel^{ZZ} = - C_{\phi\phi} \,, \qquad
   F_\parallel^{WW} = - C_{\phi\phi} \,.
\end{equation}
The fact that these form factors are given in terms of the Wilson coefficient of the operator containing two scalar fields is a nice expression of the Goldstone-boson equivalence theorem \cite{Cornwall:1974km,Vayonakis:1976vz,Chanowitz:1985hj}. The remaining longitudinal form factors vanish. 

From (\ref{VVampl}) we see that the $S\to V_1 V_2$ decay amplitudes scale like $M$ and hence are of ${\cal O}(\lambda^0)$ in SCET power counting. The corresponding decay rates can be obtained from the general expression
\begin{equation}\label{SVVrate}
\begin{aligned}
   \Gamma(S\to V_1 V_2) &= S_{V_1 V_2}\,\frac{M^2}{16\pi M_S}\,\lambda^{1/2}(x_1,x_2)
    \left[ 2 \left( |F_\perp^{V_1 V_2}|^2 + |\widetilde F_\perp^{V_1 V_2}|^2 \right)
     + |F_\parallel^{V_1 V_2}|^2 \right] ,
\end{aligned}
\end{equation}
where $x_i\equiv m_i^2/M_S^2$, and $\lambda(x,y)=(1-x-y)^2-4xy$. The factor $S_{V_1 V_2}$ takes into account a symmetry factor 1/2 for identical bosons and a color factor $(N_c^2-1)=8$ for the digluon rate. By measuring the polarizations of the vector bosons it would be possible to separately probe the three form factors characterizing each decay.

\subsection[Effective Lagrangian at ${\cal O}(\lambda^3)$]{\boldmath Effective Lagrangian at ${\cal O}(\lambda^3)$}
\label{subsec:3.2}

The operators arising at subleading order in the expansion in $\lambda$ contain fermion fields. We decompose Dirac matrices appearing in bilinears of the form $\bar\X_{n_1}\dots\X_{n_2}$ as
\begin{equation}\label{gammadecomp}
   \gamma^\mu = \frac{\nsl_1}{n_1\cdot n_2}\,n_2^\mu + \frac{\nsl_2}{n_1\cdot n_2}\,n_1^\mu
    + \gamma_\perp^\mu \,,
\end{equation}
such that $n_1{}_\mu\gamma_\perp^\mu=n_2{}_\mu\gamma_\perp^\mu=0$. Pulling out a factor $1/M$ to make the Wilson coefficients dimensionless, we find that the most general effective Lagrangian can be written in the form 
\begin{equation}\label{Leff3}
\begin{aligned}
   {\cal L}_{\rm eff}^{(3)}
   &= \frac{1}{M}\,\bigg[ C_{F_L\bar f_R}^{\,ij}(M_S,M,\mu)\,O_{F_L\bar f_R}^{\,ij}(\mu)
    + \!\sum_{k=1,2} \int_0^1\!du\,C_{F_L\bar f_R\,\phi}^{(k)\,ij}(u,M_S,M,\mu)\,
    O_{F_L\bar f_R\,\phi}^{(k)\,ij}(u,\mu) + \mbox{h.c.} \bigg] \\
   &\quad\mbox{}+ \frac{1}{M}\!\sum_{A=G,W,B} \bigg[ 
    \int_0^1\!du\,C_{F_L\bar F_L A}^{\,ij}(u,M_S,M,\mu)\,O_{F_L\bar F_L A}^{\,ij}(u,\mu) 
    + (F_L\to f_R) + \mbox{h.c.} \bigg] \,,
\end{aligned}
\end{equation} 
where we have defined the mixed-chirality operators
\begin{equation}\label{O35}
\begin{aligned}
   O_{F_L\bar f_R}^{\,ij}(\mu)
   &= S\,\bar\X_{L,n_1}^{\,i} \Phi_0\,\X_{R,n_2}^{\,j} + (n_1\leftrightarrow n_2) \,, \\
   O_{F_L\bar f_R\,\phi}^{(1)\,ij}(u,\mu)
   &= S\,\bar\X_{L,n_1}^{\,i} \Phi_{n_1}^{(u)}\,\X_{R,n_2}^{\,j} + (n_1\leftrightarrow n_2) \,, \\
   O_{F_L\bar f_R\,\phi}^{(2)\,ij}(u,\mu)
   &= S\,\bar\X_{L,n_1}^{\,i} \Phi_{n_2}^{(u)}\,\X_{R,n_2}^{\,j} + (n_1\leftrightarrow n_2) \,,
\end{aligned}
\end{equation} 
and the same-chirality operators
\begin{equation}\label{O36}
\begin{aligned}
   O_{F_L\bar F_L A}^{\,ij}(u,\mu)
   &= S\,\bar\X_{L,n_1}^{\,i}\,\Asl_{n_1}^{\perp(u)}\,\X_{L,n_2}^{\,j}
    + (n_1\leftrightarrow n_2) \,, \\
   O_{f_R\bar f_R A}^{\,ij}(u,\mu)
   &= S\,\bar\X_{R,n_1}^{\,i}\,\Asl_{n_1}^{\perp(u)}\,\X_{R,n_2}^{\,j}
    + (n_1\leftrightarrow n_2) \,.
\end{aligned}
\end{equation} 
In (\ref{Leff3}) a sum over the flavor indices $i,j$ is implied. We do not show color and $SU(2)_L$ indices. The left-handed fermions $F_L$ are $SU(2)_L$ doublets, while the right-handed fermions $f_R$ are singlets. If the right-handed fermion field in (\ref{O35}) refers to an up-type quark, the scalar doublet $\Phi$ needs to be replaced by $\tilde\Phi$ with $\tilde\Phi_a=\epsilon_{ab}\,\Phi_b^*=(\phi_2^*,-\phi_1^*)^T$ to ensure gauge invariance. Our notation is such that, e.g., the coefficient $C_{F_L\bar f_R}^{\,ij}$ multiplies an operator which produces a left-handed fermion doublet $F_L$ with generation index $i$ and a right-handed anti-fermion $\bar f_R$ with generation index $j$. Note that, in general, the Wilson coefficients can be arbitrary complex matrices in generation space. 

When SCET operators contain two or more collinear fields belonging to the same jet, the total collinear momentum $P_i$ carried by the jet is shared by the various particles described by these fields. Each component field carries a positive fraction $u_j$ of the large component $\bar n_i\cdot P_i$, such that $\sum_j u_j=1$. The product of Wilson coefficients times operators then becomes generalized to a convolution in these variables. In our discussion above a single variable $u$ appears, which refers to the longitudinal momentum fraction carried by the boson field. To see how it arises, consider the first operator in (\ref{O36}) as an example. Its contribution to the effective Lagrangian can be written in the form (leaving out flavor indices and omitting a second term with $n_1\leftrightarrow n_2$ for simplicity)
\begin{equation}\label{example3}
\begin{aligned}
   &\int\!dr\,ds\,dt\,\bar C_{F_L\bar F_L A}(r,s,t,M,\mu)\,
    S(x)\,\bar\X_{L,n_1}(x+s\bar n_1)\,\Asl_{n_1}^\perp\big(x+(r+s)\bar n_1\big)\,
    \X_{L,n_2}(x+t\bar n_2) \\
   &= \int\!dr\,C_{F_L\bar F_L A}(r,\bar n_1\cdot{\cal P}_1,\bar n_2\cdot{\cal P}_2,M,\mu)\,
    S(x)\,\bar\X_{L,n_1}(x)\,\Asl_{n_1}^\perp(x+r\bar n_1)\,\X_{L,n_2}(x) \,,
\end{aligned}
\end{equation} 
where the Wilson coefficient in the second step is defined in analogy with (\ref{FTC}). To complete the switch to momentum space we take a Fourier transform of the Wilson coefficient with respect to $r$. This gives
\begin{equation}
\begin{aligned}
   &\int\!d\omega\,
    C_{F_L\bar F_L A}(\omega,\bar n_1\cdot{\cal P}_1,\bar n_2\cdot{\cal P}_2,M,\mu)
    \int\frac{dr}{2\pi}\,e^{-i\omega r}\,
    S(x)\,\bar\X_{L,n_1}(x)\,\Asl_{n_1}^\perp(x+r\bar n_1)\,\X_{L,n_2}(x) \\    
   &= \int\!d\omega\,
    C_{F_L\bar F_L A}(\omega,\bar n_1\cdot{\cal P}_1,\bar n_2\cdot{\cal P}_2,M,\mu)\,  
    S(x)\,\bar\X_{L,n_1}(x) \left[ \delta(i\bar n_1\cdot\partial+\omega)\,
    \Asl_{n_1}^\perp(x) \right] \X_{L,n_2}(x) \,.
\end{aligned}
\end{equation} 
The $\delta$-function ensures that the variable $\omega$ is set equal to the the large (outgoing) momentum component $\bar n_1\cdot p_A$ carried by $n_1$-collinear gauge field. Since this must be a fraction of the large component $\bar n_1\cdot P_1$ of the total collinear momentum, it is useful to replace $\omega=u\,\bar n_1\cdot{\cal P}_1$ in the final step. This yields
\begin{equation}
\begin{aligned}
   & \int\!du\,C_{F_L\bar F_L A}(u\,\bar n_1\cdot{\cal P}_1,\bar n_1\cdot{\cal P}_1,
    \bar n_2\cdot{\cal P}_2,M,\mu)\,
    \delta\bigg(u-\frac{\bar n_1\cdot{\cal P}_1^{A}}{\bar n_1\cdot{\cal P}_1}\bigg)\,
    S(x)\,\bar\X_{L,n_1}(x)\,\Asl_{n_1}^\perp(x)\,\X_{L,n_2}(x) \,.
\end{aligned}
\end{equation} 
The operator $\bar n_1\cdot{\cal P}_1^{A}$ picks out the large momentum component carried by the gauge field, whereas $\bar n_1\cdot{\cal P}_1$ produces the large momentum component carried by all $n_1$-collinear fields together. Using reparameterization invariance, the Wilson coefficient in this expression can be rewritten in the form $C_{F_L\bar F_L A}(u,M_S,M,\mu)$
shown in (\ref{Leff3}), where we also use the short-hand notation
\begin{equation}
   S\,\bar\X_{L,n_1}^{\,i}\,\Asl_{n_1}^{\perp(u)}\,\X_{L,n_2}^{\,j} 
   \equiv \delta\bigg(u-\frac{\bar n_1\cdot{\cal P}_1^{A}}{\bar n_1\cdot{\cal P}_1}\bigg)\,
    S(x)\,\bar\X_{L,n_1}(x)\,\Asl_{n_1}^\perp(x)\,\X_{L,n_2}(x) \,.
\end{equation} 

Several additional comments are in order. First, we do not include same-chirality operators in (\ref{O36}) in which instead of $\Asl_{n_1}^\perp$ there is a derivative $i\delsl_\perp$ acting on one of the collinear building blocks. These operators can be reduced to those in (\ref{Leff3}) using the equations of motion. For instance, one finds that 
\begin{equation}
\begin{aligned}
   S\,\bar\X_{L,n_1}^{\,i} i\delsl_\perp \X_{L,n_2}^{\,j} + (n_1\leftrightarrow n_2)
   &= \!\int_0^1\!du\,\bigg[ \left( \bm{Y}_f \right)^{jk}\!\left( O_{F_L\bar f_R}^{ik} 
    + O_{F_L\bar f_R\,\phi}^{(2)\,ik} \right)
    - \sum_r\,\big( O_{F_L\bar F_L A_r}^{ji} \big)^\dagger \bigg] \,, \\
   S\,\big[ \bar\X_{L,n_1}^{\,i} (-i\!\overleftarrow{\delsl}\hspace{-1mm}_\perp)\,\X_{L,n_2}^{\,j} 
    \big] + (n_1\leftrightarrow n_2)
   &= \!\int_0^1\!du\,\bigg[ \big( \bm{Y}_f^* \big)^{ik}\!\left[ \big( O_{F_L\bar f_R}^{jk} \big)^\dagger 
    + \big( O_{F_L\bar f_R\,\phi}^{(2)\,jk} \big)^\dagger \right]
    - \sum_r\,O_{F_L\bar F_L A_r}^{ij} \bigg] \,,
\end{aligned}
\end{equation}
where $\bm{Y}_f$ with $f=u,d,e$ are the SM Yukawa matrices (for quarks the expressions on the right-hand side must be summed over $f=u,d$), and the sums over $r$ run over the different gauge bosons which couple to the fermion described by $\X_L$. Similar relations hold for the corresponding operators involving right-handed fields. Secondly, in addition to the operators in (\ref{O36}), one can construct operators in which the indices of the transverse objects $\A_{n_1\perp}^\mu$ and $\gamma_\perp^\nu$ are contracted using the $\epsilon_{\mu\nu}^\perp$ tensor defined in (\ref{gepsdef}). However, these operators can be reduced to those in (\ref{O36}) using the identity (with $\gamma_5=i\gamma^0\gamma^1\gamma^2\gamma^3$)
\begin{equation}
   [\gamma_\mu^\perp, \gamma_\nu^\perp]
   = - i\epsilon_{\mu\nu}^\perp\,\frac{[\nsl_1,\nsl_2]}{n_1\cdot n_2}\,\gamma_5 \,,
\end{equation}
which holds in four spacetime dimensions \cite{Hill:2004if}.\footnote{In dimensional regularization, so-called ``evanescent'' operators containing anti-symmetric products of more than two $\gamma_\perp^\mu$ matrices can appear at loop order. A regularization scheme including the effects of these operators must be employed for higher-order calculations. This is the two-dimensional analogue, in the space of transverse directions, of the standard procedure employed in four dimensions \cite{Buras:1989xd,Dugan:1990df}.} 
From this relation it follows that 
\begin{equation}
   P_{n_1}^\dagger \epsilon_{\mu\nu}^\perp\,\gamma_\perp^\nu P_{n_2}
   = i P_{n_1}^\dagger \gamma_\mu^\perp\gamma_5\,P_{n_2} \,.
\end{equation}

Finally, we note that at ${\cal O}(\lambda^3)$ there do not appear operators containing two collinear fermion fields belonging to the same jet. These operators would need to include the bilinears (modulo $L\leftrightarrow R$)
\begin{equation}
   \bar\X_{L,n_1}^{\,i} \frac{\nbsl_1}{2}\,\X_{L,n_1}^{\,j} = {\cal O}(\lambda^2) 
    \qquad \mbox{or} \qquad
   \bar\X_{L,n_1}^{\,i} \Phi_{n_i} \frac{\nbsl_1}{2}\,\gamma_\mu^\perp\,\X_{R,n_1}^{\,j} 
    = {\cal O}(\lambda^3) \,,
\end{equation}
where $\gamma_\mu^\perp$ is now defined with respect to the plane spanned by the vectors $n_1$ and $\bar n_1$, and the subscript $n_i$ on the scalar doublet could be 0, $n_1$ or $n_2$. In case of the first operator, the required $n_2$-collinear field could be $n_2\cdot\A_{n_2}$, $\Phi_{n_2}^\dagger\Phi_{n_2}$, $(\Phi_{n_2}^\dagger\Phi_{n_1}+\mbox{h.c.})$, or $(\Phi_{n_2}^\dagger\Phi_0+\mbox{h.c.})$, all of which are of ${\cal O}(\lambda^2)$. In the second case, the open Lorentz index must be contracted with $\A_{n_i\perp }^\mu$ or $\partial_\perp^\mu$, both of which count as ${\cal O}(\lambda)$. Hence, any such operator is at least of ${\cal O}(\lambda^4)$.

The effective Lagrangian (\ref{Leff3}) describes the two-body decays of $S$ into a pair of SM fermions. Taking into account that external collinear fermions have power counting $\lambda^{-1}$, it follows that the $S\to f\bar f$ decay amplitudes obey the scaling rule shown in (\ref{scalings}). At tree level, only the operator $O_{F_L\bar f_R}$ and its hermitian conjugate give non-zero contributions. After electroweak symmetry breaking the fermion fields must be rotated from the weak to the mass basis, and in the process the Wilson coefficients in (\ref{Leff3}), which are matrices in generation space, are transformed as well. In matrix notation, we have e.g.\
\begin{equation}
   \bm{C}_{F_L\bar f_R}\to \bm{U}_{f_L}^\dagger \bm{C}_{F_L\bar f_R} \bm{W}_{f_R} 
   \equiv {\bf C}_{f_L\bar f_R} \,,
\end{equation}
where $f_L$ (with a lower case) now refers to one of the two members of the left-handed doublet, and $\bm{U}_{f_L}$ and $\bm{W}_{f_R}$ with $f=u,d,e$ denote the rotation matrices transforming the left-handed and right handed fermions from the weak to the mass basis. In order not to clutter our notation too much, we use the same symbol but with a straight ``C'' instead of the slanted ``$C$'' for the Wilson coefficients in the mass basis. We then find the non-zero decay amplitudes 
\begin{equation}\label{eq41}
\begin{aligned}
   {\cal M}(S\to f_{iL}\,\bar f_{jR})
   &= \frac{v}{\sqrt2 M}\,{\rm C}_{f_L\bar f_R}^{\,ij} 
    \bar u_L(k_1) P_{n_1}^\dagger P_{n_2} v_R(k_2) 
    = \frac{v}{\sqrt2}\,\frac{M_S}{M}\,{\rm C}_{f_L\bar f_R}^{\,ij}\,e^{i\varphi_{ij}}\,, \\
   {\cal M}(S\to f_{iR}\,\bar f_{jL})
   &= \frac{v}{\sqrt2\,M}\,{\rm C}_{f_L\bar f_R}^{\,ji\,*} 
    \bar u_R(k_1) P_{n_1}^\dagger P_{n_2} v_L(k_2) 
    = \frac{v}{\sqrt2}\,\frac{M_S}{M}\,{\rm C}_{f_L\bar f_R}^{\,ji\,*}\,e^{-i\varphi_{ji}} \,,
\end{aligned}
\end{equation}
where $i,j$ are flavor indices. Note that the products of two highly energetic fermion spinors give rise to the appearance of the hard scale $M_S$ in the matrix elements of the SCET operators. The expressions on the right hold up to some complex phases, which depend on the phase conventions for the fermion fields. The corresponding decay rates are given by (with $x_i=m_i^2/M_S^2$)
\begin{equation}
\begin{aligned}
   \Gamma(S\to f_{iL}\,\bar f_{jR})
   &= N_c^f\,\frac{v^2 M_S}{32\pi M^2}\,\lambda^{1/2}(x_i,x_j)\,
    \big| {\rm C}_{f_L\bar f_R}^{\,ij} \big|^2 \,, \\
   \Gamma(S\to f_{iR}\,\bar f_{jL})
   &= N_c^f\,\frac{v^2 M_S}{32\pi M^2}\,\lambda^{1/2}(x_i,x_j)\,
    \big| {\rm C}_{f_L\bar f_R}^{\,ji} \big|^2 \,,
\end{aligned}
\end{equation}
where $N_c^f$ is a color factor, which equals~3 for quarks and 1 for leptons. Beyond the Born approximation, the remaining operators in (\ref{Leff3}) also contribute to the decay rates. In Section~\ref{sec:RGEs} we will study the mixing of these operators under renormalization. 

In general, the couplings of $S$ to fermions contain both CP-even and CP-odd terms. Let us decompose the various complex matrices of Wilson coefficients in the mass basis into their real and imaginary components, for example
\begin{equation}
   {\bf C}_{f_L\bar f_R}
   \equiv {\bf K}_{f_L\bar f_R} + i\hspace{0.3mm}\widetilde{\bf K}_{f_L\bar f_R} \,,
\end{equation}
and likewise for ${\bf C}_{f_L\bar f_R\,\phi}^{(i)}$ and ${\bf C}_{f_L\bar f_R A}$. Under a CP transformation the effective Lagrangian (\ref{Leff3}) transforms into an analogous expression with all Wilson coefficients replaced by their complex conjugates. It follows that the terms involving the real parts of the coefficients (${\bf K}_{f_L\bar f_R}$ etc.) are CP even, while those involving the imaginary parts ($\widetilde{\bf K}_{f_L\bar f_R}$ etc.) are CP odd.

\subsection[Effective Lagrangian at ${\cal O}(\lambda^4)$]{\boldmath Effective Lagrangian at ${\cal O}(\lambda^4)$}
\label{subsec:phi4}

The only two-body decay of the heavy resonance $S$ not yet accounted for is $S\to Zh$. Operators mediating this decay arise first at NNLO in the $\lambda$ expansion. At this order a large number of new operators arise, but only a single operator contributes to the $S\to Zh$ decay amplitude at tree level. It reads
\begin{equation}\label{Leff4}
\begin{aligned}
   {\cal L}_{\rm eff}^{(4)} 
   &\ni \frac{\widetilde C_{\phi\phi\phi\phi}(M_S,M,\mu)}{M}
    \Big[ iS\! \left( \Phi_{n_1}^\dagger \Phi_0 - \Phi_0^\dagger\,\Phi_{n_1} \right)\!
    \left( \Phi_{n_2}^\dagger \Phi_0 + \Phi_0^\dagger\,\Phi_{n_2} \right) 
    + (n_1\leftrightarrow n_2) \Big] \\
   &= \frac{\widetilde C_{\phi\phi\phi\phi}(M_S,M,\mu)}{M}\,2iS
    \left( \Phi_{n_1}^\dagger \Phi_0\,\Phi_{n_2}^\dagger \Phi_0
     - \Phi_0^\dagger\,\Phi_{n_1}\,\Phi_0^\dagger\,\Phi_{n_2} \right) .
\end{aligned}
\end{equation}
The tilde on the Wilson coefficient indicates that this operator is CP odd \cite{Bauer:2016zfj}. The corresponding decay amplitude is given by
\begin{equation}
   {\cal M}(S\to Zh) = -i\widetilde C_{\phi\phi\phi\phi}\,\frac{v^2 m_Z}{M}\,
    \frac{\bar n_1\cdot\varepsilon_\parallel^*(k_1)}{\bar n_1\cdot k_1} \,.
\end{equation}
It vanishes unless the $Z$ boson is longitudinally polarized, in which case one finds
\begin{equation}
   {\cal M}(S\to Z_\parallel h) = -i\widetilde C_{\phi\phi\phi\phi}\,\frac{v^2}{M} \,,
\end{equation}
in accordance with (\ref{scalings}). To derive this result, we have used the exact representation
\begin{equation}
   \varepsilon_\parallel^\mu(k_1)
   = \frac{k_1\cdot k_2}{m_1\,\big[(k_1\cdot k_2)^2 - m_1^2\,m_2^2\big]^{1/2}}
    \left( k_1^\mu - \frac{m_1^2}{k_1\cdot k_2}\,k_2^\mu \right)
\end{equation}
for the longitudinal polarization vector. For the decay rate, we obtain (with $x_i=m_i^2/M_S^2$)
\begin{equation}
   \Gamma(S\to Zh) = \frac{v^4}{16\pi M_S M^2}\,\lambda^{1/2}(x_Z,x_h)\,
    \big| \widetilde C_{\phi\phi\phi\phi} \big|^2 \,.
\end{equation}
The puzzling fact that the $S\to Zh$ decay amplitude scales like $\lambda^2$, whereas all other diboson amplitudes scale like $\lambda^0$, finds a natural explanation in our approach.

The complete list of the operators arising at ${\cal O}(\lambda^4)$ in the effective Lagrangian describing the two-body decays of the heavy resonance $S$ is rather extensive. It includes operators containing $S$ along with four scalar fields, four transverse gauge fields, two scalar fields and two transverse gauge fields, four fermion fields, two fermion fields and two scalar/transverse gauge fields, and two fermion fields and an ultra-soft gauge or scalar field. Moreover, in some of these operators a transverse gauge field can be replaced by a transverse derivative, or two transverse gauge fields can be replaced by a small component of a collinear gauge field or an ultra-soft gauge field. A complete classification of these operators is left for future work.

\section{\boldmath SCET$_{\rm BSM}$ for three-body decays of $S$}
\label{sec:3body}

The construction of the effective Lagrangian describing three-body decays of the heavy resonance $S$ proceeds in analogy with Section~\ref{sec:SCET}. Generically, the three SM particles in the final state have momenta aligned with three different directions $\bm{n}_i$ with $i=1,2,3$, and hence the scalar products $k_i\cdot k_j={\cal O}(M_S^2)$ are set by the mass scale of the decaying particle. The leading SCET operators involving three $n_i$-collinear fields are of ${\cal O}(\lambda^3)$ and contain fermion bilinears. The corresponding operators can be constructed as in Section~\ref{subsec:3.2}. The purely bosonic three-body decays $S\to hhh$, $S\to h V_1 V_2$ and $S\to V_1 V_2 V_3$ appear first at ${\cal O}(\lambda^4)$ in the SCET expansion. They will not be considered in detail here.

Without loss of generality, we choose the outgoing boson along the direction $\bm{n}_3$. Dirac matrices are still decomposed as shown in (\ref{gammadecomp}), where now $n_1\cdot n_2=1-\cos\phi_{12}$ with $\phi_{12}=\arcangle(\bm{n_1},\bm{n}_2)$ is no longer equal to~2. We find 
\begin{equation}\label{Leff3body}
\begin{aligned}
   {\cal L}_{\rm eff}^{(3)}
   &= \frac{1}{M} \left[ D_{F_L\bar f_R\,\phi}^{\,ij}(\{m_{kl}^2\},M,\mu)\,
    Q_{F_L\bar f_R\,\phi}^{\,ij}(\mu) + \mbox{h.c.} \right] \\
   &\quad\mbox{}+ \frac{1}{M} \sum_{A=G,W,B} \Big[ 
    D_{F_L\bar F_L A}^{\,ij}(\{m_{kl}^2\},M,\mu)\,Q_{F_L\bar F_L A}^{\,ij}(\mu) 
    + D_{f_R\bar f_R A}^{\,ij}(\{m_{kl}^2\},M,\mu)\,Q_{f_R\bar f_R A}^{\,ij}(\mu) \Big] \,,
\end{aligned}
\end{equation} 
with
\begin{equation}\label{O35b}
\begin{aligned}
   Q_{F_L\bar f_R\,\phi}^{\,ij}(\mu)
   &= S\,\bar\X_{L,n_1}^{\,i} \Phi_{n_3} \X_{R,n_2}^{\,j} + (n_1\leftrightarrow n_2) \,, \\
   Q_{F_L\bar F_L A}^{\,ij}(\mu)
   &= S\,\bar\X_{L,n_1}^{\,i} \gamma_\mu^\perp\A_{n_3\perp}^\mu\,\X_{L,n_2}^{\,j} 
    + (n_1\leftrightarrow n_2) \,, \\
   Q_{f_R\bar f_R A}^{\,ij}(\mu)
   &= S\,\bar\X_{R,n_1}^{\,i} \gamma_\mu^\perp\A_{n_3\perp}^\mu\,\X_{R,n_2}^{\,j} 
    + (n_1\leftrightarrow n_2) \,.
\end{aligned}
\end{equation}
Once again $i$, $j$ are flavor indices. Note that the symbol $\perp$ on $\gamma_\mu^\perp$ means ``perpendicular to the plane spanned by $n_1$ and $n_2$'', see (\ref{gammadecomp}), while on the gauge field $\A_{n_3\perp}^\mu$ it means ``perpendicular to the plane spanned by $n_3$ and $\bar n_3$'', see (\ref{Aperpdef}). The contraction of these two objects gives rise to a non-trivial dependence on the light-like reference vectors of the three final-state particles, shown in relation (\ref{ndots}) below.

We denote the Wilson coefficients by $D$ and the operators by $Q$ in order to distinguish them from the corresponding quantities in the Lagrangian for two-body decays shown in (\ref{Leff3}). If the right-handed fermion field in (\ref{O35b}) refers to an up-type quark, the scalar doublet $\Phi_{n_3}$ needs to be replaced by $\tilde\Phi_{n_3}$ to ensure gauge invariance. The Wilson coefficients $\bm{D}_{F_L\bar f_R\,\phi}$ are arbitrary complex matrices in generation space, while $\bm{D}_{F_L\bar F_L A}$ and $\bm{D}_{f_R\bar f_R A}$ are hermitian matrices. As before, we will denote the corresponding coefficients after transformation to the mass basis with an unslanted symbol ``D'' (and use $f_L$ instead of $F_L$ to represent one of the two members of the weak doublet).

Note that there are no convolution integrals in (\ref{Leff3body}), in contrast with (\ref{Leff3}). On the other hand, by a generalization of the argument given before (\ref{eq16}), the Wilson coefficients can now depend on the three invariants (with $k\ne l\in\{1,2,3\}$)
\begin{equation}
   \frac{n_k\cdot n_l}{2}\,\bar n_k\cdot{\cal P}_k\,\bar n_l\cdot{\cal P}_l
   = \left( \frac{n_k}{2}\,\bar n_k\cdot{\cal P}_k + \frac{n_l}{2}\,\bar n_l\cdot{\cal P}_l
    \right)^2 \simeq ({\cal P}_k + {\cal P}_l)^2 \,. 
\end{equation}
For a three-body decay, these invariants evaluate to the squared invariant masses $m_{kl}^2$ of the different pairs of final-state particles, which are subject to the relation
\begin{equation}
   m_{12}^2 + m_{23}^2 + m_{13}^2
   = M_S^2 + m_1^2 + m_2^2 + m_3^2 \simeq M_S^2 \,.
\end{equation}

It is straightforward to derive from (\ref{Leff3body}) the relevant tree-level expressions for the 3-body decay amplitudes of the heavy resonance $S$. Since both the Wilson coefficients and the matrix elements of the effective Lagrangian depend on the pair invariant masses squared, we can only compute the doubly differential decay rate, summed over polarizations of the vector boson where appropriate, in two of these variables (the so-called Dalitz-plot distribution) in a model-independent way. 

We begin with the decay modes mediated by the opposite-chirality operators in (\ref{Leff3body}), for which we obtain
\begin{equation}\label{rate1}
   \frac{d^2\Gamma(S\to f_{iL}\,\bar f_{jR}\,h)}{dm_{12}^2\,dm_{23}^2}
   = \frac{d^2\Gamma(S\to f_{iL}\,\bar f_{jR}\,Z)}{dm_{12}^2\,dm_{23}^2}
   = \frac{N_c^f}{512\pi^3 M_S^3}\,\frac{m_{12}^2}{M^2}\,
    \big| {\rm D}_{f_L\bar f_R\,\phi}^{ij} \big|^2 \,,
\end{equation}    
and     
\begin{equation}\label{rate2}
   \frac{d^2\Gamma(S\to f_{iL}\,\bar f_{jR}\,W^\pm)}{dm_{12}^2\,dm_{23}^2}
   = \frac{N_c^f}{256\pi^3 M_S^3}\,\frac{m_{12}^2}{M^2}\,
    \big| {\rm D}_{f_L\bar f_R\,\phi}^{ij} \big|^2 \,,
\end{equation}
where as before $N_c^f=3$ for quarks and~1 for leptons. Here $m_{12}^2=m_{f\bar f}^2$ and $m_{23}^2=m_{\bar f h}^2$ or $m_{\bar f V}^2$. Analogous expressions hold with $L\leftrightarrow R$ on the left-hand side and $i\leftrightarrow j$ on the right-hand side. To arrive at these results, we have used that 
\begin{equation}
   \frac{n_1\cdot n_2}{2}\,\bar n_1\cdot k_1\,\bar n_2\cdot k_2
   \simeq 2k_1\cdot k_2 \simeq m_{12}^2 \,.
\end{equation}
Only the longitudinal polarization state of the electroweak gauge bosons contributes to these rates.

\begin{table}
\centering
\begin{tabular}{l|cc}
\hline
Process & Color/coupling factor & Coefficient \\
\hline
$S\to f_{iL}\,\bar f_{jL}\,\gamma$ & $N_c^f\alpha$
 & $T_3^{f_L}\,{\rm D}_{f_L\bar f_L W}^{ij} + Y_{f_L} {\rm D}_{f_L\bar f_L B}^{ij}$ 
 \\[1mm]
$S\to f_{iR}\,\bar f_{jR}\,\gamma$ & $N_c^f\alpha$
 & $Y_{f_R} {\rm D}_{f_R\bar f_R B}^{ij}$ \\[1mm]
\hline 
$S\to f_{iL}\,\bar f_{jL}\,Z$ & $N_c^f\alpha$
 & $T_3^{f_L}\,\frac{c_w}{s_w}\,{\rm D}_{f_L\bar f_L W}^{ij} 
     - \frac{s_w}{c_w}\,Y_{f_L} {\rm D}_{f_L\bar f_L B}^{ij}$ \\[1mm]
$S\to f_{iR}\,\bar f_{jR}\,Z$ & $N_c^f\alpha$
 & $- \frac{s_w}{c_w}\,Y_{f_R} {\rm D}_{f_R\bar f_R B}^{ij}$ \\[1mm]
\hline 
$S\to f_{iL}\,\bar f_{jL}\,W^\pm$ & $N_c^f\alpha$
 & $\frac{1}{s_w}\,{\rm D}_{f_L\bar f_L W}^{ij}$ \\[1mm]
$S\to f_{iR}\,\bar f_{jR}\,W^\pm$ & $N_c^f\alpha$
 & $0$ \\[1mm]
\hline
$S\to q_{iL}\,\bar q_{jL}\,g$ & $N_c C_F \alpha_s$
 & ${\rm D}_{f_L\bar f_L G}^{ij}$ \\[1mm]
$S\to q_{iR}\,\bar q_{jR}\,g$ & $N_c C_F \alpha_s$
 & ${\rm D}_{f_R\bar f_R G}^{ij}$ \\[1mm]
\hline
\end{tabular}
\caption{\label{tab:charges}
Color factors, gauge couplings and Wilson coefficients entering the expressions for the doubly differential decay rates for the three-body decays $S\to f_{iL}\,\bar f_{jL}\,V$ and $S\to f_{iR}\,\bar f_{jR}\,V$, all of which are given by a formula analogous to (\ref{beauty}).}
\end{table}

From the same-chirality operators in (\ref{Leff3body}) we obtain slightly more complicated expressions. Focusing on the case where a fermion pair is produced along with a photon, we find
\begin{equation}\label{beauty}
\begin{aligned}
   \frac{d^2\Gamma(S\to f_{iL}\,\bar f_{jL}\,\gamma)}{dm_{12}^2\,dm_{23}^2}
   &= \frac{N_c^f\alpha}{32\pi^2 M_S^3}\,\frac{m_{12}^2}{M^2}\,
    \frac{(m_{13}^2)^2 + (m_{23}^2)^2}{(M_S^2-m_{12}^2)^2}
    \left| T_3^{f_L}\,{\rm D}_{f_L\bar f_L W}^{ij}
    + Y_{f_L} {\rm D}_{f_L\bar f_L B}^{ij} \right|^2 , \\
   \frac{d^2\Gamma(S\to f_{iR}\,\bar f_{jR}\,\gamma)}{dm_{12}^2\,dm_{23}^2}
   &= \frac{N_c^f\alpha}{32\pi^2 M_S^3}\,\frac{m_{12}^2}{M^2}\,
    \frac{(m_{13}^2)^2 + (m_{23}^2)^2}{(M_S^2-m_{12}^2)^2}
    \left| Y_{f_R} {\rm D}_{f_R\bar f_R B}^{ij} \right|^2 ,
\end{aligned}
\end{equation}
where $T_3^{f_L}$ denotes the weak isospin of the left-handed fermion, and $Y_{f_L}$, $Y_{f_R}$ are the hypercharges of the fermions. Only the two transverse polarization states of the vector bosons contribute to these rates. The squared decay amplitudes depend in a non-trivial way on the light-like reference vectors of the final-state mesons. We find that they involve the quantity
\begin{equation}\label{ndots}
   \frac{n_1\cdot n_3\,n_2\cdot\bar n_3+n_2\cdot n_3\,n_1\cdot\bar n_3}{n_1\cdot n_2}
   \simeq 2\,\frac{(m_{13}^2)^2 + (m_{23}^2)^2}{(M_S^2-m_{12}^2)^2} \,.
\end{equation}
To derive this result, we have replaced $n_i\cdot\bar n_3=2n_i\cdot v-n_i\cdot n_3$, where $v^\mu$ is the 4-velocity of the decaying resonance $S$. We have then multiplied all light-like vectors with the corresponding energies (defined in the rest frame of $S$) to obtain $k_i^\mu\simeq E_i n_i^\mu$, and at the end eliminated the energies using that $m_{12}^2=(k_1+k_2)^2=(M_S v-k_3)^2\simeq M_S^2-2M_S E_3$ etc. The decay rates for the production of fermion pairs along with other gauge bosons are given by analogous expressions with different charge and color factors and involving different combinations of Wilson coefficients, as shown in Table~\ref{tab:charges}.

Neglecting the masses of the final-state particles, the boundaries of the Dalitz plot are such that
\begin{equation}
   0 < m_{12}^2 < M_S^2 \,, \qquad
   0 < m_{23}^2 < M_S^2 - m_{12}^2 \,.
\end{equation}
Since our results have been derived under the assumption that the invariant mass of each pair of final-state particles is of order $M_S$, strictly speaking they are not valid near the boundary of the Dalitz plot. On the other hand, since the boundary effect occurs in a power-suppressed region of phase space, one usually does not need to worry about this issue, unless the squared decay amplitude is singular near the boundary. 

If the Wilson coefficients only depend on $m_{12}^2$ but not on $m_{23}^2$ and $m_{13}^2$ individually, the expressions in (\ref{rate1}), (\ref{rate2}) and (\ref{beauty}) can be integrated over $m_{23}^2$ to obtain the distributions in the invariant mass of the fermion pair. We will show in Section~\ref{sec:hierarchy} that this condition is satisfied (at least at tree level) in all models featuring a double hierarchy $M\gg M_S\gg v$. We quote the result for the interesting case of the decay $S\to t\bar t Z$. Summing over the different polarization states of the fermions, and defining $x_{12}=m_{t\bar t}^2/M_S^2$, we find
\begin{equation}
\begin{aligned}
   \frac{d\Gamma(S\to t\bar tZ)}{dx_{12}}
   &= \frac{N_c M_S^3}{512\pi^3 M^2}\,x_{12}(1-x_{12})\,\bigg\{
    \Big[ \big| {\rm D}_{u_L\bar u_R\,\phi}^{33}(x_{12}) \big|^2 
    + \big| {\rm D}_{u_R\bar u_L\phi}^{33}(x_{12}) \big|^2 \Big] \\
   &\quad\mbox{}+ \frac{32\pi\alpha}{3}\,
    \bigg[ \left| \frac{c_w}{2s_w}\,{\rm D}_{u_L\bar u_L W}^{33}(x_{12}) 
     - \frac{s_w}{6c_w}\,{\rm D}_{u_L\bar u_L B}^{33}(x_{12}) \right|^2 
     + \left| \frac{2s_w}{3c_w}\,{\rm D}_{u_R\bar u_R B}^{33}(x_{12}) \right|^2 \bigg] \bigg\} .
\end{aligned}
\end{equation}
With the help of (\ref{rate1}), (\ref{rate2}) and Table~\ref{tab:charges}, all other rates can be obtained from this expression by means of simple substitutions.

\section{Evolution equations for the Wilson coefficients}
\label{sec:RGEs}

Large logarithms of the scale ratio $M_S/v$ can be systematically resummed to all orders in perturbation theory using our effective theory. The leading effects arise from Sudakov double logarithms related to the interplay of soft and collinear emissions of virtual particles. They are controlled by so-called cusp logarithms in the anomalous dimensions of SCET operators \cite{Bauer:2000yr}, which govern the scale dependence of the Wilson coefficients in the effective Lagrangian of SCET$_{\rm BSM}$. The relevant anomalous dimensions are computed from the UV divergences of SCET operators and are independent of the masses of the SM particles. They can be most conveniently derived by setting all masses to zero and using off-shell external momenta as infrared regulators. The relevant version of the effective theory is called SCET$_{\rm I}$. It describes the interactions of $n_i$-collinear fields with so-called ultra-soft fields with momentum scaling $(\lambda^2,\lambda^2,\lambda^2)$ \cite{Bauer:2001yt,Beneke:2002ph}. Note that the ultra-soft scale $\lambda^2 M_S\sim v^2/M_S$ lies parametrically below the characteristic scale $v$ of the low-energy theory. This scale arises in intermediate steps of the calculation, but it drops out from the final expressions for the anomalous dimensions.\footnote{It would be possible to calculate the anomalous dimensions using the masses of the SM particles as infrared regulators. In this case the ultra-soft scale does not arise (except in graphs involving massless gauge-boson exchange), but the calculations are far more complicated due to the appearance of rapidity divergences, which require analytic regulators beyond dimensional regularization \cite{Becher:2010tm,Chiu:2011qc}.}

The discussion in this section is considerably more technical than that in previous sections. The reader not interested in these technicalities may directly proceed with Section~\ref{sec:hierarchy}, noting however that there is a well-defined formalism which allows us to derive the evolution equations needed to resum large logarithms in the SCET$_{\rm BSM}$.

\subsection{Operators containing a single field in each collinear direction}

The scale dependence of the Wilson coefficients of operators containing a single $n_i$-collinear field for each direction of large energy flow can be described by a universal anomalous dimension depending on scalar products formed out of the different collinear momenta $\{\underline{p}\}=\{p_1,\dots,p_n\}$ (strictly speaking the momenta $p_i$ should be replaced by the corresponding label operators ${\cal P}_i$), such that \cite{Becher:2009cu}
\begin{equation}\label{RGevol}
   \mu\,\frac{d}{d\mu}\,C(\{\underline{p}\},\mu)
   = \Gamma(\{\underline{p}\},\mu)\,C(\{\underline{p}\},\mu) \,.
\end{equation}
For the Wilson coefficients of operators containing at most three external particles, the all-order structure of the anomalous dimension is extremely simple: It contains so-called ``dipole terms'' for pairs of particles $i$ and $j$, which involve logarithms of the kinematic invariants $s_{ij}=2p_i\cdot p_j$ (with all momenta outgoing) and correlations of the two particles in the space of group generators, as well as single-particle terms for each field \cite{Becher:2009cu,Gardi:2009qi,Becher:2009qa,Dixon:2009ur}. Moreover, using charge conservation, one can eliminate all group generators in terms of the eigenvalues of the quadratic Casimir operators $C_i\in\{C_F,C_A\}$ for particles transforming in the fundamental or the adjoint representation of the gauge group. The two-particle terms involve the universal cusp anomalous dimension for light-like Wilson loops \cite{Korchemsky:1987wg}. Since the SM gauge group is a direct product of three simple groups $G_r$ with $G_1=U(1)_Y$, $G_2=SU(2)_L$ and $G_3=SU(3)_c$, the cusp terms involve a sum over the three group factors. The anomalous dimensions for two- and three-particle operators take the form
\begin{equation}\label{Gamma}
\begin{aligned}
   \Gamma(\{p_1,p_2\},\mu) 
   &= \sum_r C_1^{(r)}\,\gamma_{\rm cusp}^{(r)}\,\ln\frac{-s_{12}-i0}{\mu^2}
    + \sum_{i=1,2}\,\gamma^i \,, \\
   \Gamma(\{p_1,p_2,p_3\},\mu) 
   &= \frac12\,\sum_r \sum_{\pi(i,j,k)} \left( C_i^{(r)} + C_j^{(r)} - C_k^{(r)} \right)
    \gamma_{\rm cusp}^{(r)}\,\ln\frac{-s_{ij}-i0}{\mu^2}
    + \sum_{i=1,2,3}\,\gamma^i \,,
\end{aligned}
\end{equation}
where $\pi(i,j,k)$ refers to the even permutations of $(1,2,3)$. For non-abelian $SU(N)$ groups one has $C_F^{(r)}=(N^2-1)/(2N)$ and $C_A^{(r)}=N$. For the hypercharge group $G_1=U(1)_Y$ one sets $C_F^{(1)}=Y_i^2$ and $C_A^{(1)}=0$, where $Y_i$ denotes the hypercharge of the particle $i$. If a particle does not transform under a group $G_r$, then $C_i^{(r)}$ is set to zero. 

The single-particle anomalous dimensions $\gamma^i$ for fermions contain terms involving the SM Yukawa matrices, which multiply the Wilson coefficients in (\ref{RGevol}) from the left (for a field $\bar\X$ producing an outgoing fermion) or from the right (for a field $\X$ producing an outgoing anti-fermion).

From (\ref{Gamma}), it is straightforward to derive exact all-order relations for the anomalous dimensions governing the scale dependence of the Wilson coefficients of the two-jet operators in the effective Lagrangian (\ref{Leff2}) arising at ${\cal O}(\lambda^2)$ and for the three-jet operators in the effective Lagrangian (\ref{Leff3body}) arising at ${\cal O}(\lambda^3)$. Omitting all arguments for simplicity, we obtain
\begin{equation}\label{GammaGG}
\begin{aligned}
   \Gamma_{\phi\phi}
   &= \left( \frac14\,\gamma_{\rm cusp}^{(1)} + \frac34\,\gamma_{\rm cusp}^{(2)} \right) 
    \bigg( \ln\frac{M_S^2}{\mu^2} - i\pi \bigg) + 2\gamma^\phi \,, \\
   \Gamma_{BB} &= \widetilde\Gamma_{BB} = 2\gamma^B \,, \\
   \Gamma_{WW} &= \widetilde\Gamma_{WW}
    = 2\gamma_{\rm cusp}^{(2)}\,\bigg( \ln\frac{M_S^2}{\mu^2} - i\pi \bigg) + 2\gamma^W \,, \\
   \Gamma_{GG} &= \widetilde\Gamma_{GG}
    = 3\gamma_{\rm cusp}^{(3)}\,\bigg( \ln\frac{M_S^2}{\mu^2} - i\pi \bigg) + 2\gamma^G \,,
\end{aligned}   
\end{equation}
and
\begin{equation}\label{Gamma3body}
\begin{aligned}
   \Gamma_{F_L\bar f_R\,\phi}^Q 
   &= \left[ \frac12 \left( Y_{F_L}^2 + Y_{f_R}^2 - Y_\phi^2 \right) \gamma_{\rm cusp}^{(1)} 
    + \delta_{fq}\,\frac43\,\gamma_{\rm cusp}^{(3)} \right] 
    \bigg( \ln\frac{m_{12}^2}{\mu^2} - i\pi \bigg) \\
   &\quad\mbox{}+ \left[ \frac12 \left( Y_\phi^2 + Y_{F_L}^2 - Y_{f_R}^2 \right)
    \gamma_{\rm cusp}^{(1)} + \frac34\,\gamma_{\rm cusp}^{(2)} \right]
    \bigg( \ln\frac{m_{13}^2}{\mu^2} - i\pi \bigg) \\
   &\quad\mbox{}+ \frac12 \left( Y_\phi^2 + Y_{f_R}^2 - Y_{F_L}^2 \right)
    \gamma_{\rm cusp}^{(1)}\,\bigg( \ln\frac{m_{23}^2}{\mu^2} - i\pi \bigg)
    + \gamma^{F_L} + \gamma^{\bar f_R} + \gamma^\phi \,, \\
   \Gamma_{f_R\bar F_L\phi}^Q
   &= \Gamma_{F_L\bar f_R\,\phi}^Q(m_{13}^2\leftrightarrow m_{23}^2,F_L\leftrightarrow f_R) \,, \\
   \Gamma_{F_L\bar F_L B}^Q 
   &= \left[ Y_{F_L}^2 \gamma_{\rm cusp}^{(1)} + \frac34\,\gamma_{\rm cusp}^{(2)} 
    + \delta_{fq}\,\frac43\,\gamma_{\rm cusp}^{(3)} \right] 
    \bigg( \ln\frac{m_{12}^2}{\mu^2} - i\pi \bigg) 
    + \gamma^{F_L} + \gamma^{\bar F_L} + \gamma^B \,, \\
   \Gamma_{F_L\bar F_L W}^Q
   &= \left[ Y_{F_L}^2 \gamma_{\rm cusp}^{(1)} - \frac14\,\gamma_{\rm cusp}^{(2)} 
    + \delta_{fq}\,\frac43\,\gamma_{\rm cusp}^{(3)} \right] 
    \bigg( \ln\frac{m_{12}^2}{\mu^2} - i\pi \bigg) \\
   &\quad\mbox{}+ \gamma_{\rm cusp}^{(2)}\,
    \bigg( \ln\frac{m_{13}^2}{\mu^2} + \ln\frac{m_{23}^2}{\mu^2} - 2i\pi \bigg) 
    + \gamma^{F_L} + \gamma^{\bar F_L} + \gamma^W \,, \\
   \Gamma_{Q_L\bar Q_L G}^Q 
   &= \left[ Y_{Q_L}^2 \gamma_{\rm cusp}^{(1)} + \frac34\,\gamma_{\rm cusp}^{(2)} 
    - \frac16\,\gamma_{\rm cusp}^{(3)} \right] \bigg( \ln\frac{m_{12}^2}{\mu^2} - i\pi \bigg) \\
   &\quad\mbox{}+ \frac32\,\gamma_{\rm cusp}^{(3)}\,
    \bigg( \ln\frac{m_{13}^2}{\mu^2} + \ln\frac{m_{23}^2}{\mu^2} - 2i\pi \bigg) 
    + \gamma^{Q_L} + \gamma^{\bar Q_L} + \gamma^G \,, \\
   \Gamma_{f_R\bar f_R B}^Q 
   &= \left[ Y_{f_R}^2 \gamma_{\rm cusp}^{(1)} 
    + \delta_{fq}\,\frac43\,\gamma_{\rm cusp}^{(3)} \right] 
    \bigg( \ln\frac{m_{12}^2}{\mu^2} - i\pi \bigg) 
    + \gamma^{f_R} + \gamma^{\bar f_R} + \gamma^B \,, \\
   \Gamma_{q_R\bar q_R G}^Q 
   &= \left[ Y_{q_R}^2 \gamma_{\rm cusp}^{(1)} - \frac16\,\gamma_{\rm cusp}^{(3)} \right] 
    \bigg( \ln\frac{m_{12}^2}{\mu^2} - i\pi \bigg) \\
   &\quad\mbox{}+ \frac32\,\gamma_{\rm cusp}^{(3)}\,
    \bigg( \ln\frac{m_{13}^2}{\mu^2} + \ln\frac{m_{23}^2}{\mu^2} - 2i\pi \bigg) 
    + \gamma^{q_R} + \gamma^{\bar q_R} + \gamma^G \,,
\end{aligned}   
\end{equation}
where $\delta_{fq}=1$ if the fermion is a quark and 0 otherwise. We have indicated the anomalous dimensions of the three-jet operators by a superscript ``Q''.

In general, the cusp anomalous dimensions $\gamma_{\rm cusp}^{(r)}$ and the single-particle anomalous dimensions $\gamma^i$ depend on the three gauge couplings $\alpha_1=\alpha/c_w^2$, $\alpha_2=\alpha/s_w^2$ and $\alpha_3=\alpha_s$, the quartic scalar coupling, and the Yukawa couplings. Up to two-loop order, however, the cusp anomalous dimension for the gauge group $G_r$ only depends on the corresponding coupling $\alpha_r$. Explicitly, it is given by \cite{Korchemsky:1987wg,Korchemskaya:1992je,Jantzen:2005az}
\begin{equation}
   \gamma_{\rm cusp}^{(r)}
   = \frac{\alpha_r}{\pi} + \left[ \left( \frac{67}{36} - \frac{\pi^2}{12} \right) C_A^{(r)} 
    - \sum_f\,\frac{5}{18}\,T_F^{(r)} d_f - \frac{1}{9}\,T_F^{(r)} d_\phi \right] 
    \left( \frac{\alpha_r}{\pi} \right)^2 + \dots \,,
\end{equation}
where $T_F^{(r)}=1/2$ for the non-abelian groups ($r=2,3$) and $T_F^{(1)}=Y_i^2$ for the hypercharge group. The coefficients $d_f$ and $d_\phi$ are the dimensions of the representations of the chiral fermions and the scalar doublet with respect to the other two gauge groups. The sum runs over the chiral fermion multiplets of the SM model, and we have used that there is a single complex scalar doublet.\footnote{In the same notation, the one-loop coefficient of the $\beta$ function for a given gauge coupling reads \cite{Gross:1973ju}
\[
   \beta_0^{(r)} 
   = \frac{11}{3}\,C_A^{(r)} - \sum_f\,\frac23\,T_F^{(r)} d_f - \frac13\,T_F^{(r)} d_\phi \,.
\]
}
Explicitly, one finds
\begin{equation}\label{Gamcusp}
\begin{aligned}
   \gamma_{\rm cusp}^{(1)} 
   &= \frac{\alpha_1}{\pi} - \frac{17}{6} \left( \frac{\alpha_1}{\pi} \right)^2 + \dots \,, \qquad 
   \gamma_{\rm cusp}^{(2)}
   = \frac{\alpha_2}{\pi} + \left( 2 - \frac{\pi^2}{6} \right)
    \left( \frac{\alpha_2}{\pi} \right)^2 + \dots \,, \\ 
   \gamma_{\rm cusp}^{(3)} 
   &= \frac{\alpha_3}{\pi} + \left( \frac{47}{12} - \frac{\pi^2}{4} \right)
    \left( \frac{\alpha_3}{\pi} \right)^2 + \dots \,.
\end{aligned}   
\end{equation}
The three-loop coefficient of the cusp anomalous dimension is only known for a single gauge group and neglecting the contributions from the scalar Higgs doublet \cite{Moch:2004pa}.

We will restrict our discussion here to a consistent resummation of Sudakov logarithms at leading logarithmic order. This requires the calculation of the cusp anomalous dimension to two-loop order, as given in (\ref{Gamcusp}), while the remaining anomalous dimensions are required with one-loop accuracy. For fermions and the scalar doublet, the one-loop coefficients from gauge interactions in units of $\alpha_r/\pi$ are $-3C_F^{(r)}/4$ \cite{Becher:2009qa} and $-C_F^{(r)}$ \cite{Chiu:2007yn}, respectively. The one-loop coefficients of the anomalous dimensions of the gauge fields vanish, since in contrast to \cite{Becher:2009qa} we have included the gauge couplings in the definitions of the $n_i$-collinear gauge fields in (\ref{calAdef}). Including also the contributions from the Yukawa interactions to the wave-function renormalizations of the fields, we obtain
\begin{equation}
\begin{aligned}
   \gamma^{f_L} = \gamma^{\bar f_L} 
   &=- Y_{f_L}^2\,\frac{\alpha_1}{4\pi} - \frac{9\alpha_2}{16\pi} 
    - \delta_{fq}\,\frac{\alpha_3}{\pi} 
    + \frac{1}{32\pi^2}\,\bm{Y}_f\bm{Y}_f^\dagger \,, \\
   \gamma^{f_R} = \gamma^{\bar f_R} 
   &= - Y_{f_R}^2\,\frac{\alpha_1}{4\pi} - \delta_{fq}\,\frac{\alpha_3}{\pi} 
    + \frac{1}{16\pi^2}\,\bm{Y}_f^\dagger\bm{Y}_f \,, \\
   \gamma^\phi &= - \frac{\alpha_1}{4\pi} - \frac{3\alpha_2}{4\pi} 
    + \sum_f\,\frac{N_c^f y_f^2}{16\pi^2} \,,
\end{aligned}   
\end{equation}
where in the last expression the sum runs over the different fermion species, and $y_f$ denotes the Yukawa coupling of the fermion $f$.

\subsection[Two-jet operators at ${\cal O}(\lambda^3)$]{\boldmath Two-jet operators at ${\cal O}(\lambda^3)$}
\label{subsec:5.2}

For operators containing more than one $n_i$-collinear field in a given direction, the anomalous dimensions are more complicated than the simple expressions shown in (\ref{Gamma}). This concerns, in particular, the anomalous dimensions governing the scale dependence of the Wilson coefficients of the two-jet operators arising at ${\cal O}(\lambda^3)$ in the SCET$_{\rm BSM}$ Lagrangian, which we have defined in (\ref{O35}) and (\ref{O36}). Since these operators depend on a variable $u$ (the fraction of the total collinear momentum carried by the boson field), the anomalous dimensions are distribution-valued functions. Also, there is a non-trivial mixing of these operators under renormalization. Finally, we will find that some of the convolution integrals appearing in the evolution equations exhibit endpoint singularities at the boundary of the integration domain, which need to be treated with care. For simplicity, we will only explore the effects of QCD evolution here, leaving a more complete treatment to future work. We will thus assume that the fermion fields in the three-jet operators are quark fields.

The presence of the scalar doublet implies that, as far as QCD evolution is concerned, the mixed-chirality operators in (\ref{O35}) renormalize like two-jet operators, with anomalous dimensions given by (in this section we keep the dependence on the color factors $C_F=4/3$ and $C_A=3$ explicit)
\begin{equation}\label{eq73}
\begin{aligned}
   \Gamma_{Q_L\bar q_R} 
   &= C_F\,\gamma_{\rm cusp}^{(3)}\,\bigg( \ln\frac{M_S^2}{\mu^2} - i \pi \bigg)
    + 2\gamma^q \,, \\
   \Gamma_{Q_L\bar q_R\,\phi}^{(i)} 
   &= C_F\,\gamma_{\rm cusp}^{(3)}\,\bigg( \ln\frac{(1-u) M_S^2}{\mu^2} - i\pi \bigg)
    + 2\gamma^q \,; \quad i=1,2 \,,
\end{aligned}   
\end{equation}
where we have used that $\gamma^{Q_L}=\gamma^{q_R}\equiv\gamma^q=-3C_F\alpha_s/(4\pi)+\dots$ under QCD evolution. The same is true for the same-chirality operators for which the gauge field belongs to $SU(2)_L$ or $U(1)_Y$, i.e.\
\begin{equation}
   \Gamma_{Q_L\bar Q_L B} = \Gamma_{q_R\bar q_R B} = \Gamma_{Q_L\bar Q_L W} 
   = \Gamma_{q_R\bar q_R W}
   = C_F\,\gamma_{\rm cusp}^{(3)}\,\bigg( \ln\frac{(1-u)M_S^2}{\mu^2} - i\pi \bigg)
    + 2\gamma^q \,.
\end{equation}
When only QCD corrections are taken into account, the cusp anomalous dimension \cite{Moch:2004pa} and the anomalous dimension of the quark field \cite{Moch:2005id,Becher:2006mr} are known to three-loop order. 

\begin{figure}
\begin{center}
\includegraphics[width=0.19\textwidth]{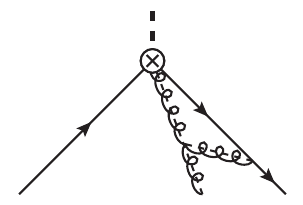} 
\includegraphics[width=0.19\textwidth]{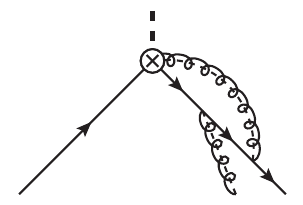} 
\includegraphics[width=0.19\textwidth]{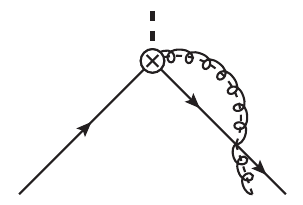} 
\includegraphics[width=0.19\textwidth]{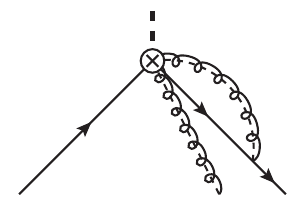} \\ 
\includegraphics[width=0.19\textwidth]{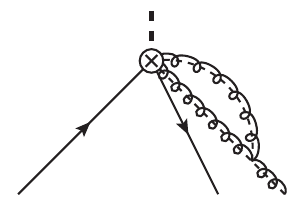}
\includegraphics[width=0.19\textwidth]{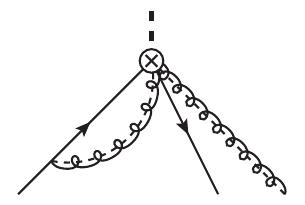}
\includegraphics[width=0.19\textwidth]{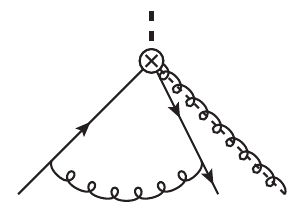} 
\includegraphics[width=0.19\textwidth]{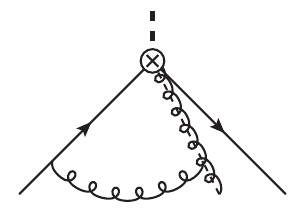} 
\end{center}
\vspace{-2mm}
\caption{\label{fig:Gammagraphs} 
One-loop diagrams contributing to the anomalous dimension $\Gamma_{q\bar q G}$ in (\ref{Gam55}). The short dashed line represents the heavy scalar resonance $S$. Solid lines denote collinear quarks, curly lines with dashes denote collinear gluons, and simple curly lines represent ultra-soft gluons. Collinear fields moving along the same direction are drawn next to each other.}
\end{figure}

The same-chirality operators containing a gluon field exhibit a more interesting behavior. Due to the dependence of the operators $O_{Q_L\bar Q_L G}$ and $O_{q_R\bar q_R G}$ on the variable $u$, the anomalous dimension governing the multiplicative renormalization of these operators is a distribution-valued function of two variables $u$ and $w$. We find that the scale dependence of the corresponding Wilson coefficients is determined by the evolution equation (with $q=Q_L$ or $q_R$)
\begin{equation}\label{Gam55}
   \mu\,\frac{d}{d\mu}\,\bm{C}_{q\bar q G}(u,M_S,M,\mu) 
   = \int_0^1\!dw\,\Gamma_{q\bar q G}(u,w,M_S,\mu)\,\bm{C}_{q\bar q G}(w,M_S,M,\mu) \,,
\end{equation}
where here and below we use a boldface notation to indicate that the Wilson coefficients are matrices in generation space. The anomalous dimension $\Gamma_{q\bar q G}$ can be calculated in analogy with the derivation of the anomalous dimensions of the subleading SCET current operators arising in $B$-meson physics performed in \cite{Hill:2004if,Beneke:2005gs} (see \cite{Beneke:2017ztn} for related recent work). It is convenient to use the background-field gauge \cite{Abbott:1980hw} for the external gluon, in which the combination $g_s\,G^{\mu,a}$ is not renormalized. Evaluating the UV divergences of the one-loop diagrams shown in Figure~\ref{fig:Gammagraphs}, supplemented by wave-function renormalization, we obtain (with $\bar u\equiv 1-u$ and $\bar w\equiv 1-w$)
\begin{equation}\label{superGamma}
\begin{aligned}
   \Gamma_{q\bar q G}(u,w,M_S,\mu) 
   &= \left[ C_F \left( \ln\frac{\bar u M_S^2}{\mu^2} - i \pi - \frac32 \right) 
    + \frac{C_A}{2} \left( \ln\frac{u}{\bar u} + 1 \right) \right] 
    \gamma_{\rm cusp}^{(3)}\,\delta(u-w) \\
   &\quad\mbox{}+ \bar w\,\Big[ V_1(\bar u,\bar w) + V_2(\bar u,\bar w) \Big] 
    + {\cal O}(\alpha_s^2) \,.
\end{aligned}
\end{equation}
The logarithmic terms in the first line are exact to all orders in perturbation theory, whereas the remaining terms have been computed at one-loop order. The kernel functions $V_i$, which are symmetric in their arguments, have been computed first in \cite{Hill:2004if}. At one-loop order one finds 
\begin{equation}
\begin{aligned}
   V_1(\bar u,\bar w) + V_2(\bar u,\bar w) 
   &= - \frac{C_A}{2}\,\frac{\alpha_s}{\pi}\,\Bigg\{ \frac{1}{\bar u\bar w}
    \left[ \bar u\,\frac{\theta(u-w)}{u-w} + \bar w\,\frac{\theta(w-u)}{w-u} \right]_+ \\
   &\hspace{24mm}\mbox{}+ \left( \frac{w}{\bar w} - \frac{1}{u} \right) \theta(u-w)
    + \left( \frac{u}{\bar u} - \frac{1}{w} \right) \theta(w-u) \Bigg\} \\
   &\quad\mbox{}+ \left( C_F - \frac{C_A}{2} \right) \frac{\alpha_s}{\pi} 
    \left[ \left( 2 - \frac{\bar u\bar w}{uw} \right) \theta(u+w-1)
    + \frac{uw}{\bar u\bar w}\,\theta(1-u-w) \right] ,
\end{aligned}
\end{equation}
where for symmetric functions $g(u,w)$ the plus distribution is defined to act on test functions $f(w)$ as
\begin{equation}
   \int_0^1\!dw \left[ g(u,w) \right]_+ f(w) = \int_0^1\!dw\,g(u,w) \left[ f(w) - f(u) \right] .
\end{equation}
Using arguments based on conformal symmetry, it was shown in \cite{Hill:2004if} how the convolution in (\ref{Gam55}) can be diagonalized by expanding the Wilson coefficients in a suitable basis of Jacobi polynomials. This will be discussed in more detail elsewhere. 

\begin{figure}
\begin{center}
\includegraphics[width=0.2\textwidth]{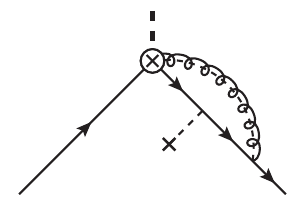} 
\includegraphics[width=0.2\textwidth]{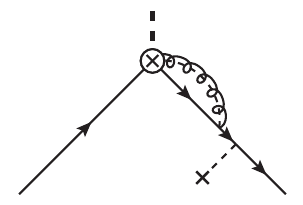} 
\includegraphics[width=0.2\textwidth]{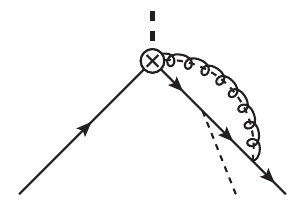} 
\includegraphics[width=0.2\textwidth]{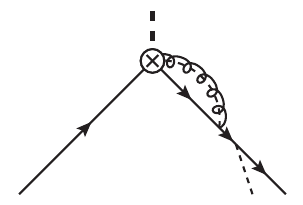} 
\includegraphics[width=0.2\textwidth]{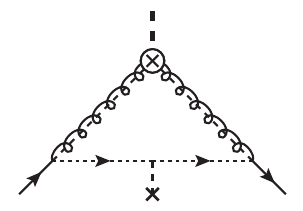} 
\end{center}
\vspace{-2mm}
\caption{\label{fig:opmix} 
{\sc Top:} One-loop diagrams responsible for the mixing of the operators $O_{Q_L\bar Q_L G}$ and $O_{q_R\bar q_R G}$ into the three mixed-chirality operators in (\ref{O35}). A dashed line ended by a cross indicates a zero-momentum scalar field $\Phi_0$, while a dashed line bending to the right shows a collinear scalar field. {\sc Bottom:} Mixing of the ${\cal O}(\lambda^2)$ operators $O_{GG}$ and $\widetilde O_{GG}$ into the operator $O_{Q_L\bar q_R}$ by means of subleading interactions in the SCET Lagrangian. The dotted line represents an ultra-soft quark.}
\end{figure}

Next, we find that the operators $O_{Q_L\bar Q_L G}$ and $O_{q_R\bar q_R G}$ mix into the three mixed-chirality operators in (\ref{O35}). The diagrams responsible for this mixing are shown in the top row of Figure~\ref{fig:opmix}. The evolution equations for the Wilson coefficients of these operators read
\begin{equation}\label{eq77}
\begin{aligned}
   \mu\,\frac{d}{d\mu}\,\bm{C}_{Q_L\bar q_R\,\phi}^{(1)}(u,\mu) 
   &= \Gamma_{Q_L\bar q_R\,\phi}^{(1)}(u,\mu)\,\bm{C}_{Q_L\bar q_R\,\phi}^{(1)}(u,\mu)
    + \int_0^1\!dw\,\Gamma_{\rm mix}(u,w,\mu)\,\bm{Y}_q(\mu)\,\bm{C}_{q_R\bar q_R G}(w,\mu) \,, \\
   \mu\,\frac{d}{d\mu}\,\bm{C}_{Q_L\bar q_R\,\phi}^{(2)}(u,\mu) 
   &= \Gamma_{Q_L\bar q_R\,\phi}^{(2)}(u,\mu)\,\bm{C}_{Q_L\bar q_R\,\phi}^{(2)}(u,\mu)
    + \int_0^1\!dw\,\Gamma_{\rm mix}(u,w,\mu)\,\bm{C}_{Q_L\bar Q_L G}^\dagger(w,\mu)\,
    \bm{Y}_q(\mu) \,,
\end{aligned}
\end{equation}
and (only if $C_{GG}=\widetilde C_{GG}=0$\,!)
\begin{equation}\label{eq78}
\begin{aligned}
   \mu\,\frac{d}{d\mu}\,\bm{C}_{Q_L\bar q_R}(\mu) 
   &= \Gamma_{Q_L\bar q_R}(\mu)\,\bm{C}_{Q_L\bar q_R}(\mu) \\
   &\quad\mbox{}+ \int_0^1\!dw\,\Gamma_{\rm mix}(0,w,\mu) 
    \left[ \bm{Y}_q(\mu)\,\bm{C}_{q_R\bar q_R G}(w,\mu)
    + \bm{C}_{Q_L\bar Q_L G}^\dagger(w,\mu)\,\bm{Y}_q(\mu) \right] ,
\end{aligned}
\end{equation}
where we have defined the mixing kernel
\begin{equation}\label{Gmix}
   \Gamma_{\rm mix}(u,w,\mu) 
   = \frac{C_F\alpha_s(\mu)}{\pi} \left[ \frac{\theta(1-u-w)}{1-u} - (1-w) \right] 
    + {\cal O}(\alpha_s^2) \,.
\end{equation}
The anomalous dimensions $\Gamma_{Q_L\bar q_R\,\phi}^{(i)}$ and $\Gamma_{Q_L\bar q_R}$ have been given in (\ref{eq73}). For simplicity, we have omitted the dependence of the Wilson coefficients on the new-physics scales $M_S$ and $M$, as well as the dependence of the anomalous dimensions on the scale $M_S$. 

The evolution equation (\ref{eq78}) needs to be modified if the Wilson coefficients $\bm{C}_{Q_L\bar Q_L G}(w,\mu)$ and $\bm{C}_{q_R\bar q_R G}(w,\mu)$ exhibit non-integrable singularities at the endpoint of the integration region. As we discuss in the Appendix, this happens whenever $C_{GG}\ne 0$ or $\widetilde C_{GG}\ne 0$. Hard matching contributions then produce poles in the Wilson coefficients located at $w=1$,\footnote{In higher orders of perturbation theory, the poles can be multiplied by logarithms of $(1-w)$.}  
whose residues are related to the coefficients $C_{GG}$ and $\widetilde C_{GG}$. While at first sight the presence of these poles appears to give rise to endpoint-divergent integrals of the form $\int_0^1\!dw\,\frac{1}{1-w}$ in (\ref{eq78}), a careful treatment reveals that the form of the mixing kernel in (\ref{Gmix}) must be modified in this case. The dimensionally regularized loop integral produces an extra factor $\big(w(1-w)\big)^{-\epsilon}$, which regularizes the singularities at $w=1$ at the expense of introducing a $1/\epsilon^2$ pole. Next, for $C_{GG}\ne 0$ or $\widetilde C_{GG}\ne 0$ there is an additional contribution arising from the mixing of the operators in the ${\cal O}(\lambda^2)$ effective Lagrangian (\ref{Leff3}) into the ${\cal O}(\lambda^3)$ operator $O_{Q_L\bar q_R}$, which happens via subleading terms in the SCET Lagrangian connecting collinear fields with an ultra-soft quark field. The relevant diagram is shown in the bottom row of Figure~\ref{fig:opmix}. The two effects conspire to produce an extra term in the evolution equation (\ref{eq78}) proportional to a combination of $C_{GG}$ and $\widetilde C_{GG}$ times a cusp logarithm. Details of this calculation are presented in the Appendix. The final result for the corrected form of the evolution equation (\ref{eq78}) reads
\begin{equation}\label{rgefinal}
\begin{aligned}
   \mu\,\frac{d}{d\mu}\,\bm{C}_{Q_L\bar q_R}(\mu) 
   &= \Gamma_{Q_L\bar q_R}(\mu)\,\bm{C}_{Q_L\bar q_R}(\mu) \\
   &\quad\mbox{}+ \frac{M^2}{M_S^2} \left[ \gamma_{\rm cusp}^{q\bar q}
    \left( \ln\frac{M_S^2}{\mu^2} - i\pi \right) + \tilde\gamma_{q\bar q} \right]
    g_s^2(\mu) \left( C_{GG}(\mu) + i\widetilde C_{GG}(\mu) \right) \bm{Y}_q(\mu) \\
   &\quad\mbox{}+ \int_0^1\!dw\,\Gamma_{\rm mix}(0,w,\mu) 
    \left[ \bm{Y}_q(\mu)\,\bar{\bm{C}}_{q_R\bar q_R G}(w,\mu)
    + \bar{\bm{C}}_{Q_L\bar Q_L G}^\dagger(w,\mu)\,\bm{Y}_q(\mu) \right] ,
\end{aligned}
\end{equation}
where 
\begin{equation}\label{cuspqq}
   \gamma_{\rm cusp}^{q\bar q} = \frac{C_F\alpha_s(\mu)}{\pi} + {\cal O}(\alpha_s^2) \,, \qquad
   \tilde\gamma_{q\bar q} = \frac{C_F\alpha_s(\mu)}{\pi} + {\cal O}(\alpha_s^2) \,,
\end{equation}
and the subtracted coefficients $\bar{\bm{C}}_{q\bar q G}(w,\mu)$ (with $q=Q_L$ or $q_R$) are obtained from the original ones by subtracting all terms of order $(1-w)^{-1}$ modulo logarithms. At lowest order in perturbation theory, we show in the Appendix that 
\begin{equation}
\begin{aligned}
   \bar{\bm{C}}_{Q_L\bar Q_L G}(u,\mu) 
   &= \bm{C}_{Q_L\bar Q_L G}(u,\mu) - \frac{M^2}{M_S^2}\,\frac{g_s^2(\mu)}{1-u}
    \left[ C_{GG}(\mu) - i\widetilde C_{GG}(\mu) \right] , \\
   \bar{\bm{C}}_{q_R\bar q_R G}(u,\mu) 
   &= \bm{C}_{q_R\bar q_R G}(u,\mu) - \frac{M^2}{M_S^2}\,\frac{g_s^2(\mu)}{1-u}
    \left[ C_{GG}(\mu) + i\widetilde C_{GG}(\mu) \right] .
\end{aligned}
\end{equation}
Note that the evolution equations (\ref{Gam55}) and (\ref{eq77}) do not require similar modifications, because the factor $(1-w)$ in the third line of (\ref{superGamma}) and the $\theta(1-u-w)$ function in (\ref{Gmix}) eliminate the singularities at $w=1$.

The cusp anomalous dimension $\gamma_{\rm cusp}^{q\bar q}$ in (\ref{cuspqq}) is a new object, which arises from the exchange of an ultra-soft quark between two collinear sectors. This is likely to be a new universal quantity, which arises in SCET applications beyond the leading power in the expansion parameter $\lambda$. The calculation of the two-loop coefficient of this quantity is an interesting open problem, to which we will return in future work. 

\subsection{Resummation of large logarithms}

To illustrate the results derived above, we now perform the resummation of large logarithms of the scale ratio $M_S/v$ for two representative cases, working consistently at leading logarithmic order. We focus on the examples $S\to\mbox{2~jets}$ and $S\to t\bar t+\mbox{jet}$, where in both cases the jets are seeded by gluons (quark jets contribute at subleading power only). At tree level, the expression for the $S\to\mbox{2~jets}$ rate obtained from (\ref{SVVrate}) reads
\begin{equation}
   \Gamma(S\to\mbox{2~jets}) 
   = \frac{M^2}{M_S}\,8\pi\alpha_s^2(\mu)
    \left( |C_{GG}(\mu)|^2 + |\widetilde C_{GG}(\mu)|^2 \right) .
\end{equation}
Likewise, the Dalitz distribution for the decay $S\to t\bar t+\mbox{jet}$ obtained from (\ref{beauty}) reads
\begin{equation}
   \frac{d^2\Gamma(S\to t\bar t+\mbox{jet})}{dx_{12}\,dx_{23}}
   = \frac{M_S^3}{M^2}\,\frac{\alpha_s(\mu)}{8\pi^2}\,
    \frac{x_{12} \left( x_{13}^2+x_{23}^2 \right)}{(1-x_{12})^2}
    \left( \left| {\rm D}_{u_L\bar u_L G}^{33}(\{x_{ij}\},\mu) \right|^2 
    + \left| {\rm D}_{u_R\bar u_R G}^{33}(\{x_{ij}\},\mu) \right|^2 \right),
\end{equation}
where we have defined $x_{ij}=m_{ij}^2/M_S^2$ with $x_{12}+x_{23}+x_{13}=1$. In the above relations we suppress the dependence of the Wilson coefficients on the new-physics scales $M$ and $M_S$. The scales $\mu$ on the right-hand side of the equations should be chosen equal to a characteristic scale of the process. In the first case, this should be a scale associated with the definition of the jets, while in the second case the scale should be around the top-quark mass. We will now derive how the Wilson coefficients at these low scales can be computed, at leading logarithmic order, in terms of the Wilson coefficients at the high scale $M_S$. We will focus on QCD evolution only, since this will give rise to the largest effects.

The general solution of the RG equation (\ref{RGevol}) has been derived in \cite{Becher:2006nr,Becher:2007ty}. For the specific cases considered here, where the relevant anomalous dimensions are given by $\Gamma_{GG}$ in (\ref{GammaGG}) and $\Gamma_{Q_L\bar Q_L G}^Q$, $\Gamma_{q_R\bar q_R G}^Q$ in (\ref{Gamma3body}), we obtain at leading logarithmic order
\begin{equation}
   C_{GG}(\mu) = \exp\left[ \frac{6}{49}\,g(M_S,\mu)
    + \frac67\,i\pi \ln r \right] C_{GG}(M_S) \,,
\end{equation}
with the same relation connecting $\widetilde C_{GG}(\mu)$ with $\widetilde C_{GG}(M_S)$, and
\begin{equation}
\begin{aligned}
   {\rm D}_{u_A\bar u_A G}^{33}(\{x_{ij}\},\mu) 
   &= \exp\left[ \frac{17}{147}\,g(M_S,\mu) 
    + \left( \frac47+\frac{17}{21}\,i\pi \right) \ln r \right] 
    {\rm D}_{u_A\bar u_A G}^{33}(\{x_{ij}\},M_S) \\
   &\quad\mbox{}\times \left( x_{12} \right)^{\frac{1}{21}\ln r}
    \left( x_{23}\,x_{13} \right)^{-\frac37\ln r} ,
\end{aligned}
\end{equation}
with $A=L,R$. We have defined the ratio $r=\alpha_s(\mu)/\alpha_s(M_S)$ and
\begin{equation}
   g(M_S,\mu) = \frac{4\pi}{\alpha_s(M_S)} \left( 1 - \frac{1}{r} - \ln r \right) 
    + \left( \frac{251}{21} - \pi^2 \right) \big( 1 - r + \ln r \big)
    + \frac{13}{7}\,\ln^2 r \,.
\end{equation}
These expressions apply for six massless flavors of quarks, and they should thus not be evaluated below the scale of the top-quark mass $m_t\approx 173$\,GeV. For a scalar resonance of mass $M_S=2$\,TeV, we find numerically
\begin{equation}
\begin{aligned}
   C_{GG}(m_t) 
   &\approx \left( 0.42 + 0.36\,i \right) C_{GG}(M_S) \,, \\
   {\rm D}_{u_A\bar u_A G}^{33}(\{x_{ij}\},m_t) 
   &\approx \left( 0.52 + 0.42\,i \right) \left( \frac{x_{12}^{1/9}}{x_{23}\,x_{13}} \right)^{0.11}
    {\rm D}_{u_A\bar u_A G}^{33}(\{x_{ij}\},M_S) \,,
\end{aligned}
\end{equation}
indicating that evolution effects can be quite sizable. In the second case, these effects lead to an additional, non-trivial dependence on the kinematic variables $x_{ij}$.

The solution of the RG equations governing the evolution of the Wilson coefficients of the two-jet operators arising at ${\cal O}(\lambda^3)$, which we have derived in Section~\ref{subsec:5.2}, is more complicated. These equations can either be solved by numerical integration or by constructing a suitable complete set of basis functions which diagonalize the relevant anomalous-dimension kernels \cite{Hill:2004if}. We leave a detailed discussion of these matters for future work.

\section{\boldmath SCET$_{\rm BSM}$ for the scale hierarchy $M\gg M_S\gg v$}
\label{sec:hierarchy}

While our SCET$_{\rm BSM}$ approach was designed to deal with the case where the masses of the heavy new resonance $S$ and of other, yet undiscovered new particles are of the same order, it also applies to new-physics scenarios in which there is a double hierarchy, such that $M\gg M_S\gg v$. It is interesting to study this case in some detail, as it provides a nice test case with which to illustrate our method.  

\subsection[Effective Lagrangian below the new-physics scale $M$]{\boldmath Effective Lagrangian below the new-physics scale $M$}

If the scale $M$ characterizing the new physics lies much above the scale of the resonance $S$, the undiscovered heavy particles can be integrated out in a first step, see the right panel of Figure~\ref{fig:scales}. This is the standard case of integrating out heavy virtual degrees of freedom, which are too massive to be produced as real particles. The effective Lagrangian obtained after this first step consists of local operators built out of $S$ and SM fields. We can write
\begin{equation}\label{Leffhigh}
   {\cal L}_{\rm eff}(M>\mu>M_S)
   = {\cal L}_{\rm SM} + {\cal L}_{\rm SMEFT} + {\cal L}_S \,. 
\end{equation}
Here ${\cal L}_{\rm SMEFT}$ is the EFT extension of the SM by higher-dimensional operators constructed out of SM fields only. Up to dimension-6 order, the corresponding operators have been classified in \cite{Weinberg:1979sa,Wilczek:1979hc,Buchmuller:1985jz,Leung:1984ni,Grzadkowski:2010es}. ${\cal L}_S$ describes the interactions of $S$ with itself and with SM fields. Up to dimension-5 order, we write the most general expression for this Lagrangian in the form
\begin{equation}\label{LeffS}
\begin{aligned}
   {\cal L}_S^{D\le 5}
   &= \frac12\,(\partial_\mu S)(\partial^\mu S) - V(S)
    - M\lambda_1\,S\,\phi^\dagger\phi 
    - \frac{\lambda_2}{2}\,S^2 \phi^\dagger\phi
    - \frac{\lambda_3}{6M}\,S^3 \phi^\dagger\phi
    - \frac{\lambda_4}{M}\,S \left( \phi^\dagger\phi \right)^2 \\
   &\quad\mbox{}+ \frac{c_{GG}}{M}\,\frac{\alpha_s}{4\pi}\,S\,G_{\mu\nu}^a G^{\mu\nu,a}
    + \frac{c_{WW}}{M}\,\frac{\alpha}{4\pi s_w^2}\,S\,W_{\mu\nu}^a W^{\mu\nu,a}
    + \frac{c_{BB}}{M}\,\frac{\alpha}{4\pi c_w^2}\,S B_{\mu\nu} B^{\mu\nu} \\
   &\quad\mbox{}+ \frac{\tilde c_{GG}}{M}\,\frac{\alpha_s}{4\pi}\,S\,G_{\mu\nu}^a \tilde G^{\mu\nu,a}
    + \frac{\tilde c_{WW}}{M}\,\frac{\alpha}{4\pi s_w^2}\,S\,W_{\mu\nu}^a \tilde W^{\mu\nu,a}
    + \frac{\tilde c_{BB}}{M}\,\frac{\alpha}{4\pi c_w^2}\,S B_{\mu\nu} \tilde B^{\mu\nu} \\
   &\quad\mbox{}- \frac{1}{M} \left( S\,\bar Q_L\,\hat{\bm{Y}}_u\,\tilde\phi\,u_R
    + S\,\bar Q_L\,\hat{\bm{Y}}_d\,\phi\,d_R + S\,\bar L_L\,\hat{\bm{Y}}_e\,\phi\,e_R 
    + \mbox{h.c.} \right) .
\end{aligned}
\end{equation}
Here $V(S)$ denotes the scalar potential, which in particular accounts for the mass $M_S$ of the scalar resonance. $G_{\mu\nu}^a$, $W_{\mu\nu}^a$ and $B_{\mu\nu}$ denote the field strength tensors of $SU(3)_c$, $SU(2)_L$ and $U(1)_Y$, and $\tilde G^{\mu\nu,a}=\frac12\,\epsilon^{\mu\nu\alpha\beta}\,G_{\alpha\beta}^a$ etc.\ are the dual field strengths. The quantities $\hat{\bm{Y}}_f$ with $f=u,d,e$ are arbitrary complex matrices in generation space. We have used the equations of motion for the SM fields and for the field $S$ to eliminate redundant operators, such as $S\,\phi^\dagger D^2\phi$, $(\partial^\mu S)\,(\phi^\dagger iD_\mu\phi+\mbox{h.c.})$, $(\partial^\mu S)\,\bar\psi\gamma_\mu\psi$ (with an arbitrary chiral fermion $\psi$), and $(\Box S)\,\phi^\dagger\phi$.\footnote{The authors of \cite{Franceschini:2015kwy} have used the equation of motion for the scalar Higgs doublet to eliminate the portal interaction $S\,\phi^\dagger\phi$ instead of the operator $S\,(D_\mu\phi)^\dagger(D^\mu\phi)$, which we have eliminated. This is not a suitable choice, because the portal interaction is a dimension-3 operator, whose contribution is enhanced by two powers of the cutoff scale relative to the dimension-5 operators in the effective Lagrangian.} 
Note that the coupling $M\lambda_1$ of the Higgs-portal operator $S\,\phi^\dagger\phi$ is dimensionful and naturally of order $M$ (i.e., it has a ``hierarchy problem''). Our operator basis agrees with the one obtained in \cite{Gripaios:2016xuo}, where a complete operator basis was constructed up to dimension $D=7$. Compared with \cite{Franceschini:2016gxv}, we have eliminated the redundant operator $S\,(\partial^\mu S)(\partial_\mu S)$. 

It is straightforward to calculate the tree-level contributions to the $S\to hh$, $S\to VV$ and $S\to f\bar f$ decay amplitudes from the above effective Lagrangian and to reproduce the scaling relations shown in (\ref{scalings}). The only non-trivial case concerns the $S\to Zh$ decay amplitude, for which the leading dimension-5 contribution arises at one-loop order and was calculated in \cite{Bauer:2016zfj}. The first tree-level contribution to the $S\to Zh$ decay amplitude arises from the dimension-7 operator
\begin{equation}\label{O7}
   {\cal L}_S^{D=7} \ni \frac{C_7}{M^3}\,(\partial^\mu S)\,
    (\phi^\dagger iD_\mu\phi + \mbox{h.c.})\,\phi^\dagger\phi \,.
\end{equation}
This contribution is suppressed by three powers of the new-physics scale.

\subsection[RG evolution from the new-physics scale to the scale $M_S$]{\boldmath RG evolution from the new-physics scale to the scale $M_S$}
\label{subsec:RGevol}

Up to dimension-5 order, the Wilson coefficients $\lambda_i$, $c_{VV}$, $\tilde c_{VV}$, and $\hat{\bm{Y}}_f$ in (\ref{LeffS}) evaluated at the new-physics scale $\mu_0\sim M$ encode the complete information about the UV completion of the theory at higher scales.\footnote{The Wilson coefficients of the Weinberg operators contained in ${\cal L}_{\rm SMEFT}$ also enter at this order, but they do not play a role in our analysis.} 
After these coefficients have been fixed from a matching calculation in the context of a particular model, they can be evolved from the high scale $\mu_0\sim M$ to the intermediate scale $\mu\sim M_S$ set by the mass of the resonance $S$ (see the right panel in Figure~\ref{fig:scales}). In this process, large logarithms of the scale ratio $M/M_S\gg 1$ are resummed. Since in our case $S$ is a gauge singlet under the SM, the relevant anomalous dimensions are those of the corresponding SM operators without the field $S$. For simplicity, we will consider here only the effects related to QCD evolution. 

At leading logarithmic order, only the Wilson coefficients $\hat{\bm{Y}}_f$ associated with quark fields change under scale variation, and we find (with $q=u,d$) 
\begin{equation}\label{RGEf}
   \hat{\bm{Y}}_q(\mu) 
   = \left( \frac{\alpha_s(\mu)}{\alpha_s(\mu_0)} \right)^{3C_F/\beta_0} \hat{\bm{Y}}_q(\mu_0) \,,
\end{equation}
where $\beta_0=\frac{11}{3}\,C_A-\frac23\,n_f$ is the first coefficient of the QCD $\beta$-function. All other Wilson coefficients are scale independent in this approximation. Beyond the leading order the evolution effects become more interesting. For the scale dependence of the coefficient $c_{GG}(\mu)$, which is renormalized multiplicatively, an exact solution can be written in terms of the QCD $\beta$-function \cite{Inami:1982xt,Grinstein:1988wz}. It reads  
\begin{equation}\label{cGGevol}
   c_{GG}(\mu) 
   = \frac{\beta(\alpha_s(\mu))/\alpha_s^2(\mu)}{\beta(\alpha_s(\mu_0))/\alpha_s^2(\mu_0)}\,
    c_{GG}(\mu_0) 
   = \left[ 1 + \frac{\beta_1}{\beta_0}\,\frac{\alpha_s(\mu)-\alpha_s(\mu_0)}{4\pi} + \dots \right] 
    c_{GG}(\mu_0) \,.
\end{equation}
We write the perturbative expansions of the $\beta$-function in the form 
\begin{equation}
   \frac{\beta(\alpha_s)}{\alpha_s^2} 
    = - \frac{1}{2\pi} \left( \beta_0 + \beta_1\,\frac{\alpha_s}{4\pi} + \dots \right) ,
\end{equation}
where $\beta_1=\frac{34}{3}\,C_A^2-\frac{10}{3}\,C_A n_f-2 C_F n_f$. For the CP-odd coefficient $\tilde c_{GG}(\mu)$ no exact solution is available. At NLO, one obtains
\begin{equation}\label{cGGtildeevol}
   \tilde c_{GG}(\mu) 
   = \left[ 1 + \frac{(\gamma_J^s)_1}{\beta_0}\,\frac{\alpha_s(\mu)-\alpha_s(\mu_0)}{4\pi}
    + \dots \right] \tilde c_{GG}(\mu_0) \,.
\end{equation}
Here $(\gamma_J^s)_1=-6C_F n_f$ is the two-loop coefficient in the anomalous dimension of the flavor-singlet axial-vector current \cite{Larin:1993tq}.

\begin{figure}
\begin{center}
\includegraphics[width=0.2\textwidth]{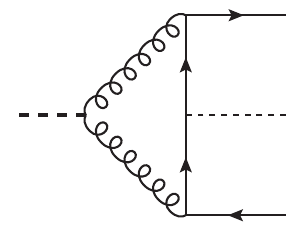} 
\end{center}
\vspace{-2mm}
\caption{\label{fig:mixinggraphs} 
One-loop diagram responsible for the mixing of the Wilson coefficients $c_{GG}$, $\tilde c_{GG}$ and $\hat{\bm{Y}}_q$ described by (\ref{RGbeauty}).}
\end{figure}

Starting at NLO, there is a non-trivial mixing of the Wilson coefficients $c_{GG}$, $\tilde c_{GG}$ and $\hat{\bm{Y}}_q$ under renormalization, caused by the diagram shown in Figure~\ref{fig:mixinggraphs}. For the CP-even, flavor-diagonal coefficients, this effect was first studied in \cite{Collins:1976yq}. Including also flavor non-diagonal couplings and CP-odd coefficients, we find that the mixing is governed by the RG equation
\begin{equation}\label{RGbeauty}
   \mu\,\frac{d}{d\mu}\,\hat{\bm{Y}}_q(\mu) 
   = \gamma^y(\mu)\,\hat{\bm{Y}}_q(\mu) 
    + \gamma^{qg}(\mu)\,\big[ c_{GG}(\mu) - i\tilde c_{GG}(\mu) \big]\,\bm{Y}_q(\mu) \,,
\end{equation}
where $\gamma^y$ is the anomalous dimension of the SM Yukawa couplings, while $\gamma^{qg}$ accounts for the mixing effects. The perturbative expansions of these objects read
\begin{equation}
   \gamma^y(\alpha_s) = \gamma_0^y\,\frac{\alpha_s}{4\pi}
    + \gamma_1^y \left( \frac{\alpha_s}{4\pi} \right)^2 + \dots \,, \qquad
   \gamma^{qg}(\alpha_s) = \gamma_1^{qg} \left( \frac{\alpha_s}{4\pi} \right)^2 + \dots \,, 
\end{equation}
where $\gamma_0^y=-6C_F$, $\gamma_1^y=-3C_F^2-\frac{97}{3}\,C_F C_A+\frac{20}{3}\,C_F T_F n_f$ \cite{Tarrach:1980up}, and $\gamma_1^{qg}=-24 C_F$. At NLO, the solution to the RG equation (\ref{RGbeauty}) takes the form
\begin{equation}\label{Yhatevol}
   \hat{\bm{Y}}_q(\mu) 
   = U_y(\mu,\mu_0) \left[ \hat{\bm{Y}}_q(\mu_0) - \frac{\gamma_1^{qg}}{2\beta_0}\,
    \frac{\alpha_s(\mu)-\alpha_s(\mu_0)}{4\pi}\, 
    \Big( c_{GG}(\mu_0) - i\tilde c_{GG}(\mu_0) \Big)\, \bm{Y}_q(\mu_0) \right] ,
\end{equation}
where
\begin{equation}
   U_y(\mu,\mu_0) 
   = \left( \frac{\alpha_s(\mu)}{\alpha_s(\mu_0)} \right)^{-\frac{\gamma_0^y}{2\beta_0}}
    \left[ 1 - \frac{\gamma_1^y\beta_0-\beta_1\gamma_0^y}{2\beta_0^2}\,
    \frac{\alpha_s(\mu)-\alpha_s(\mu_0)}{4\pi} + \dots \right] .
\end{equation}
Relations (\ref{cGGevol}), (\ref{cGGtildeevol}) and (\ref{Yhatevol}) describe the scale dependence of the Wilson coefficients between the new-physics scale $\mu_0\sim M$ and the scale $\mu\sim M_S$. 

\subsection[Matching to SCET$_{\rm BSM}$ at the scale $\mu\sim M_S$]{\boldmath Matching to SCET$_{\rm BSM}$ at the scale $\mu\sim M_S$}

At the scale $\mu\sim M_S$, the effective Lagrangian (\ref{Leffhigh}) is matched onto the SCET$_{\rm BSM}$ Lagrangians discussed in Section~\ref{sec:SCET} and~\ref{sec:3body}. The leading contributions arise from the operators of dimension up to~5. They originate from the $D=5$ operators contained in (\ref{LeffS}), or from the $D=3$ Higgs-portal interaction $S\,\phi^\dagger\phi$ in combination with a $D=6$ interaction from the effective Lagrangian ${\cal L}_{\rm SMEFT}$. We will now derive the corresponding matching conditions at tree level. In this approximation, time-ordered products of $S\,\phi^\dagger\phi$ with operators of the SMEFT Lagrangian in the basis of \cite{Grzadkowski:2010es} do not give rise to non-zero matching contributions.

\subsubsection*{\boldmath Matching coefficients at ${\cal O}(\lambda^2)$}

We begin with the Wilson coefficients of the ${\cal O}(\lambda^2)$ SCET$_{\rm BSM}$ operators in the effective Lagrangian (\ref{Leff2}), for which we obtain 
\begin{equation}\label{eq101}
\begin{aligned}
   C_{\phi\phi}(M_S,M,\mu) &= - \lambda_1 \,, && \\
   C_{GG}(M_S,M,\mu) &= - \frac{M_S^2}{M^2}\,\frac{c_{GG}}{8\pi^2} \,, 
   &\widetilde C_{GG}(M_S,M,\mu) &= \frac{M_S^2}{M^2}\,\frac{\tilde c_{GG}}{8\pi^2} \,, \\
   C_{WW}(M_S,M,\mu) &= - \frac{M_S^2}{M^2}\,\frac{c_{WW}}{8\pi^2} \,, \qquad
   &\widetilde C_{WW}(M_S,M,\mu) &= \frac{M_S^2}{M^2}\,\frac{\tilde c_{WW} }{8\pi^2} \,, \\
   C_{BB}(M_S,M,\mu) &= - \frac{M_S^2}{M^2}\,\frac{c_{BB}}{8\pi^2} \,, 
   &\widetilde C_{BB}(M_S,M,\mu) &= \frac{M_S^2}{M^2}\,\frac{\tilde c_{BB}}{8\pi^2} \,.
\end{aligned}
\end{equation}
All scale-dependent quantities are evaluated at the matching scale $\mu\sim M_S$. 

\subsubsection*{\boldmath Matching coefficients at ${\cal O}(\lambda^3)$}

\begin{figure}
\begin{center}
\includegraphics[width=0.6\textwidth]{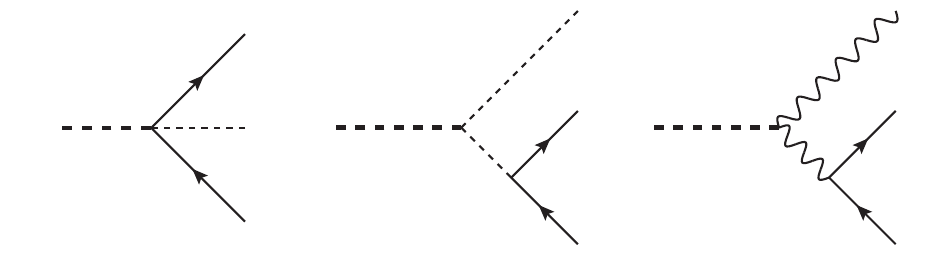} 
\end{center}
\vspace{-2mm}
\caption{\label{fig:match} 
Diagrams contributing to the tree-level matching conditions for the Wilson coefficients of ${\cal O}(\lambda^3)$ operators. The first two graphs contribute the two terms in (\ref{Cmixed3}) and (\ref{eq94}), while the third diagram generates the coefficients in (\ref{Csame3}) and (\ref{eq95}).}
\end{figure}

The matching conditions for the Wilson coefficients of the two-body ${\cal O}(\lambda^3)$ SCET$_{\rm BSM}$ operators in the effective Lagrangian (\ref{Leff3}) follow by evaluating the tree-level Feynman diagrams shown in Figure~\ref{fig:match}. We write the results in terms of matrices in generation space. For the coefficients of the mixed-chirality operators, we obtain (with $f=u,d,e$)
\begin{equation}\label{Cmixed3}
\begin{aligned}
   \bm{C}_{F_L\bar f_R}(M_S,M,\mu) 
   &= - \hat{\bm{Y}}_f - \frac{M^2\lambda_1}{M_S^2}\,\bm{Y}_f \,, \\
   \bm{C}_{F_L\bar f_R\,\phi}^{(i)}(u,M_S,M,\mu) 
   &= - \hat{\bm{Y}}_f - \frac{M^2\lambda_1}{(1-u) M_S^2}\,\bm{Y}_f \,; \quad i=1,2 \,.
\end{aligned}
\end{equation}
The matrices $\bm{Y}_f$ refer to the original Yukawa matrices of the SM. Several of the coefficients of the same-chirality operators vanish at tree level, namely 
\begin{equation}
\begin{aligned}
   \bm{C}_{L_L\bar L_L G}(u,M_S,M,\mu) &= \bm{C}_{\ell_R\bar\ell_R G}(u,M_S,M,\mu) = \bm{0} \,, \\
   \bm{C}_{f_R\bar f_R W}(u,M_S,M,\mu) &= \bm{0} \,.
\end{aligned}
\end{equation}
For the remaining coefficients, we find
\begin{equation}\label{Csame3}
\begin{aligned}
   \bm{C}_{Q_L\bar Q_L G}(u,M_S,M,\mu) &= - \frac{\alpha_s}{2\pi}\,\frac{u}{1-u}
    \left( c_{GG} + i\tilde c_{GG} \right) \bm{1} \,, \\
   \bm{C}_{q_R\bar q_R G}(u,M_S,M,\mu) &= - \frac{\alpha_s}{2\pi}\,\frac{u}{1-u}
    \left( c_{GG} - i\tilde c_{GG} \right) \bm{1} \,, \\
   \bm{C}_{F_L\bar F_L W}(u,M_S,M,\mu) 
   &= - \frac{\alpha}{2\pi s_w^2}\,\frac{u}{1-u} \left( c_{WW} + i\tilde c_{WW} \right) \bm{1} \,, \\
   \bm{C}_{F_L\bar F_L B}(u,M_S,M,\mu) 
   &= - \frac{Y_{F_L} \alpha}{2\pi c_w^2}\,\frac{u}{1-u}
    \left( c_{BB} + i\tilde c_{BB} \right) \bm{1} \,, \\
   \bm{C}_{f_R\bar f_R B}(u,M_S,M,\mu) 
   &= - \frac{Y_{f_R} \alpha}{2\pi c_w^2}\,\frac{u}{1-u}
    \left( c_{BB} - i\tilde c_{BB} \right) \bm{1} \,,
\end{aligned}
\end{equation}
where $Y_{F_L}$ and $Y_{f_R}$ in the last two relations refer to the hypercharges of the fermions. Note that at tree level these coefficients are diagonal in flavor space. Once again, all scale-dependent quantities are evaluated at the matching scale $\mu\sim M_S$.

The matching conditions for the ${\cal O}(\lambda^3)$ operators governing three-body decays of the resonance $S$ are given by similar expressions. In analogy with (\ref{Cmixed3}), we find
\begin{equation}\label{eq94}
   \bm{D}_{F_L\bar f_R\,\phi}(\{m_{kl}^2\},M,\mu) 
   = - \hat{\bm{Y}}_f - \frac{M^2\lambda_1}{m_{12}^2}\,\bm{Y}_f \,.
\end{equation}
The coefficients $\bm{D}_{F_L\bar F_L A}$ and $\bm{D}_{f_R\bar f_R A}$ are given by expressions analogous to those in (\ref{Csame3}), with the replacement $u/(1-u)\to (M_S^2-m_{12}^2)/m_{12}^2$; for example, we find
\begin{equation}\label{eq95}
   \bm{D}_{Q_L\bar Q_L G}(\{m_{kl}^2\},M,\mu) 
   = - \frac{\alpha_s}{2\pi}\,\frac{M_S^2-m_{12}^2}{m_{12}^2}
    \left( c_{GG} + i\tilde c_{GG} \right) \bm{1} \,.
\end{equation}
Note that, as anticipated in Section~\ref{sec:3body}, these results only depend on the invariant mass $m_{12}$ of the fermion pair.

The explicit expressions for the Wilson coefficients in (\ref{Cmixed3}) and (\ref{Csame3}) confirm our general arguments presented in Section~\ref{subsec:5.2}. The coefficients contain poles at $u=1$, whose residues are determined in terms of the Wilson coefficients of the ${\cal O}(\lambda^2)$ operators given in (\ref{eq101}).

\subsubsection*{\boldmath Matching coefficient $\widetilde C_{\phi\phi\phi\phi}$ at ${\cal O}(\lambda^4)$}

\begin{figure}
\begin{center}
\includegraphics[width=0.96\textwidth]{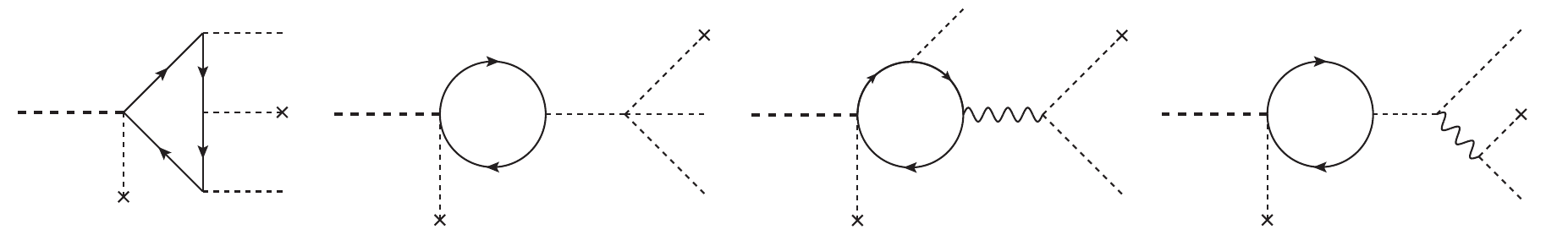} 
\end{center}
\vspace{-2mm}
\caption{\label{fig:C4matching} 
Representative one-loop diagrams contributing to the matching condition for the Wilson coefficients $\widetilde C_{\phi\phi\phi\phi}$ in (\ref{C4phi}). Dashed lines with a cross denote zero-momentum insertions of the scalar field $\varphi_0^0$. In the first and third graph one must sum over all possible attachments of the scalar lines on the fermion loop.}
\end{figure}

The coefficient $\widetilde C_{\phi\phi\phi\phi}$ in the effective Lagrangian (\ref{Leff4}) receives matching contributions starting at one-loop order. Writing the scalar doublets in the form
\begin{equation}
   \Phi_{n_i} = W_{n_i}^\dagger \left( \begin{array}{c} -i\varphi_{n_i}^+ \\ 
    \frac{1}{\sqrt2} \left( \varphi_{n_i}^0 + i\varphi_{n_i}^3 \right) \end{array} \right) , \qquad
   \Phi_0 = \frac{1}{\sqrt2} \left( \begin{array}{c} 0 \\ \varphi_0^0 \end{array} \right) ,
\end{equation}
where $\varphi_0^0$ denotes a zero-momentum boson, we find that
\begin{equation}
   {\cal L}_{\rm eff}^{(4)} 
   \ni \frac{\widetilde C_{\phi\phi\phi\phi}(M_S,M,\mu)}{M}\,S \left( \varphi_0^0 \right)^2
    \left( \varphi_{n_1}^3 \varphi_{n_2}^0 + \varphi_{n_2}^3 \varphi_{n_1}^0 \right) + \dots ,
\end{equation}
where the dots represent contributions involving more than five fields. In order to determine $\widetilde C_{\phi\phi\phi\phi}$, we compute the four-particle decay amplitude $S\to\varphi^3(k_1)\,\varphi^0(k_2)\,\varphi^0(0)\,\varphi^0(0)$ with two zero-momentum particles in the final state, in both the full theory -- defined by the Lagrangian (\ref{LeffS}) -- and the effective theory. Treating all particles other than $S$ as massless and performing the matching calculation with on-shell external states, all loop graphs in the effective theory are scaleless and hence vanish. In the full theory, the one-loop diagrams shown in Figure~\ref{fig:C4matching} give rise to non-zero results. All other diagrams are scaleless. Note that the evaluation of the two graphs involving the $B^\mu$ and $W_3^\mu$ gauge bosons requires a regulator in order to avoid that the gauge-boson propagator becomes singular. We introduce an infinitesimal momentum $q$ to the ``zero-momentum'' $\varphi^0$ boson coupling to the vector boson and take the limit $q\to 0$ after summing up all diagrams. In that way, we find in the $\overline{\rm MS}$ subtraction scheme
\begin{equation}\label{C4phi}
\begin{aligned}
   \widetilde C_{\phi\phi\phi\phi}(M_S,M,\mu)
   = - \sum_{f=u,d,e} \frac{N_c^f T_3^{f_L}}{16\pi^2}\, 
   & \bigg[ \,\mbox{Im}\,\mbox{Tr}\big( 
    \hat{\bm{Y}}_f \bm{Y}_f^\dagger \bm{Y}_f \bm{Y}_f^\dagger \big) 
    \bigg( L^2 - 2i\pi L - \frac{7\pi^2}{6} \bigg) \\
   &\hspace{2mm}\mbox{}- \mbox{Im}\,\mbox{Tr}\big(\hat{\bm{Y}}_f \bm{Y}_f^\dagger \big) 
    \bigg( 4\lambda + \frac{e^2}{2s_w^2 c_w^2} \bigg) \big( L - i\pi - 2 \big) \bigg] \,,
\end{aligned}
\end{equation}
where $\lambda$ denotes the quartic scalar coupling of the SM (not to be confused with our SCET expansion parameter), $T_3^{f_L}$ denotes the weak isospin of the left-handed fermions, and $L=\ln(M_S^2/\mu^2)$. The complex matrices $\hat{\bm{Y}}_f$ have been defined in (\ref{LeffS}), while $\bm{Y}_f$ are the Yukawa matrices of the SM. A simple result for the traces can be obtained by transforming the Yukawa matrices to the mass basis and defining 
\begin{equation}
   \big( \bm{U}_f^\dagger\,\hat{\bm{Y}}_f \bm{W}_f \big)_{ii} 
   \equiv y_{f_i} \big( c_{f_i} + i\tilde c_{f_i} \big) \,,
\end{equation}
where $y_{f_i}$ is the SM Yukawa coupling of the fermion $f_i$. This leads to
\begin{equation}
\begin{aligned}
   \widetilde C_{\phi\phi\phi\phi}(M_S,M,\mu)
   = - \sum_{f=u,d,e} \frac{N_c^f T_3^{f_L}}{16\pi^2}\,\sum_{i=1,2,3} \tilde c_{f_i}
   & \bigg[ \,y_{f_i}^4 \bigg( L^2 - 2i\pi L - \frac{7\pi^2}{6} \bigg) \\
   &\hspace{2mm}\mbox{}- y_{f_i}^2 \bigg( 4\lambda + \frac{e^2}{2s_w^2 c_w^2} \bigg) 
    \big( L - i\pi - 2 \big) \bigg] \,.
\end{aligned}
\end{equation}
The dominant contribution is likely to arise from the top quark.

In \cite{Bauer:2016zfj}, it was shown that a tree-level contribution to $\widetilde C_{\phi\phi\phi\phi}$ arises first from a dimension-7 operator in the effective Lagrangian obtained by integrating out the new-physics scale $M$, shown in (\ref{O7}). We find that the corresponding matching contribution reads 
\begin{equation}
   \delta\widetilde C_{\phi\phi\phi\phi} =  - \frac{M_S^2}{2M^2}\,C_7 \,,
\end{equation}
where $C_7$ itself is most likely suppressed by a loop factor. This contribution is parametrically suppressed compared with that in (\ref{C4phi}) by a factor $M_S^2/M^2\ll 1$.

\section{Conclusions}
\label{sec:concl}

We have developed a theoretical framework to construct a consistent effective field theory for the on-shell decays into light SM particles of the first new heavy resonance beyond the SM that will be discovered at the LHC or elsewhere. Our approach is flexible enough to retain the full dependence on the mass $M_S$ of the new resonance $S$ and on the masses of other, yet undiscovered particles. It can thus deal with the important situation where the first particle to be discovered is a member of a new sector characterized by a mass scale $M$. It provides a consistent separation between the electroweak scale $v\approx 246$\,GeV and the new-physics scales $M_S$ and $M$, irrespective of whether $M_S\sim M$ are of similar magnitude or if there is a double hierarchy $v\ll M_S\ll M$. Large double and single logarithms of scale ratios can be resummed to all orders in perturbation theory by solving RG evolution equations in the effective theory.

Our effective theory SCET$_{\rm BSM}$ is a variant of soft-collinear effective theory (SCET), in which the effective Lagrangian is constructed out of gauge-invariant collinear building blocks for the particles of the SM along with a field representing the new heavy resonance $S$. We have worked out in detail the case where $S$ is a spin-0 boson that is a singlet with respect to the SM gauge interactions. We have constructed the most general effective Lagrangian at leading, subleading, and partially subsubleading order in the expansion in $\lambda=v/M_S$. It describes all two-body decays of $S$ into SM particles. We have also constructed the leading-order effective Lagrangian describing three-body decays of $S$. We have calculated the anomalous dimensions of the operators in the effective Lagrangian and derived the RG evolution equations for their Wilson coefficients. For the operators arising at next-to-leading order in $\lambda$ several subtleties arise. These operators mix under renormalization, and their anomalous dimensions are distribution-valued functions depending on the momentum fractions carried by different collinear field operators. The evolution equations involve a new cusp anomalous dimension originating from the exchange of an ultra-soft quark between two collinear sectors. There has recently been an increasing interest in applications of SCET beyond the leading power in $\lambda$ \cite{Bonocore:2014wua,Bonocore:2015esa,Bonocore:2016awd,DelDuca:2017twk,Penin:2014msa,Liu:2017vkm,Moult:2016fqy,Boughezal:2016zws,Moult:2018jjd,Beneke:2017ztn}. The results obtained in this paper are an important contribution to this rapidly developing field.

There are several extensions and refinements of our approach which are worth pursuing. The matrix elements of the SCET$_{\rm BSM}$ operators, which we have computed at tree level, should be calculated to one-loop order. These matrix elements contain large rapidity logarithms of the scale ratio $M_S/v$ from the collinear anomaly, despite the fact that the hard scale $M_S$ has been integrated out from the low-energy effective theory. Understanding the structure of these logarithms and showing that they do not spoil factorization is an important ingredient of our approach. It will be important to complete the calculation of the one-loop anomalous dimensions of the two-jet operators arising at ${\cal O}(\lambda^3)$ in the SCET$_{\rm BSM}$ Lagrangian, which we have presented in Section~\ref{subsec:5.2}, by including the contributions from electroweak and Yukawa interactions. Perhaps more importantly, the two-loop contribution to the cusp anomalous dimension $\gamma_{\rm cusp}^{q\bar q}$ in (\ref{cuspqq}) should be calculated. This quantity is associated with the exchange of an ultra-soft quark between two collinear fields moving along different directions. It is a crucial new ingredient for a consistent Sudakov resummation at subleading power in SCET. Finally, it would be interesting to provide a complete classification of the operators arising at ${\cal O}(\lambda^4)$ in the SCET$_{\rm BSM}$ Lagrangian, whose structure we have only sketched in Section~\ref{subsec:phi4}.

Our work can be generalized in several ways. In particular, it would be interesting to extend it to other cases of new heavy resonances, which are well motivated theoretically. This includes various heavy leptoquarks or $Z'$ bosons, which have been proposed to address some present anomalies in rare and semileptonic decays of $B$ mesons \cite{Altmannshofer:2013foa,Hiller:2014yaa,Alonso:2015sja,Freytsis:2015qca,Bauer:2015knc,Barbieri:2015yvd} (see \cite{Dorsner:2016wpm} for a recent review). It also applies to heavy particles that can serve as mediators to the dark sector, generalizing the hybrid EFT framework recently proposed in \cite{Alanne:2017oqj}. Finally, it would be interesting to calculate the Wilson coefficients in the SCET$_{\rm BSM}$ Lagrangian in some concrete new-physics models. Specifically, in future work we plan to illustrate our results in the context of an extension of the SM containing heavy, vector-like fermions. 

As our community eagerly awaits the discovery of new heavy particles, we have developed here a general effective field-theory approach that allows one to describe the decays of such particles into SM particles in a model-independent way, systematically separating the new-physics scales from the scales of the SM, accounting for the full complexity of the (partially unknown) UV completion via Wilson coefficient functions and providing a framework for the resummation of large logarithms to all orders in perturbation theory.

\newpage
\subsubsection*{Acknowledgments}

We are grateful to Martin Beneke and Robert Szafron for pointing out an error in the calculation of the anomalous dimension (81) of the original version of this paper. The research of M.N.\ is supported by the Cluster of Excellence {\em Precision Physics, Fundamental Interactions and Structure of Matter\/} (PRISMA -- EXC~1098) and grant 05H12UME of the German Federal Ministry for Education and Research (BMBF). S.A.~ gratefully acknowledges support from the DFG Research Training Group {\em Symmetry Breaking in Fundamental Interactions\/} (GRK~1581). The work of M.K.~is supported by the Swiss National Science Foundation (SNF) under contract 200021-175940.

\begin{appendix}

\section{Derivation of the evolution equation (\ref{rgefinal})}
\renewcommand{\theequation}{A.\arabic{equation}}
\setcounter{equation}{0}

\subsubsection*{Effective Lagrangian of SCET}

The leading-order SCET Lagrangian describing a massless, $n$-collinear fermion (of any chirality)
\begin{equation}
   \xi_n(x) = \frac{\nsl\nbsl}{4}\,\psi(x)
\end{equation}
interacting with a (abelian or non-abelian) gauge field $A^\mu$ reads \cite{Bauer:2000yr,Beneke:2002ph}
\begin{equation}\label{LSCET}
   {\cal L}_{\xi,n}^{(0)}(x)
   = \bar\xi_n(x)\,\frac{\nbsl}{2} \left( in\cdot D + i\Dsl_{\perp c}\,\frac{1}{i\bar n\cdot D_c}\,
    i\Dsl_{\perp c} \right) \xi_n(x) + \dots \,,
\end{equation}
where the dots represent the effective Yang-Mills Lagrangian and gauge-fixing terms. The covariant collinear derivative is defined as
\begin{equation}
   iD_c^\mu = i\partial^\mu + g_A\,A_n^\mu(x) \,,
\end{equation}
where $g_A$ denotes the relevant gauge coupling. The covariant derivative without a subscript ``$c$'' is defined as
\begin{equation}
   in\cdot D = in\cdot \partial + g_A\,n\cdot A_n(x) + g_A\,n\cdot A_{us}(x_-) \,.
\end{equation}
It includes the ultra-soft gauge field $n\cdot A_{us}$ in addition to the small component of the collinear gauge field $n\cdot A_n$, both of which have the same power counting ($\sim\lambda^2$). Note that the ultra-soft gauge field is multipole-expanded and lives at position $x_-\equiv\frac{n}{2}\,\bar n\cdot x$. This ensures that only the relevant components $n\cdot p_{us}$ of ultra-soft momenta, which can compete with the corresponding small components $n\cdot p_n$ of collinear momenta, enter in the computation of Feynman diagrams. The Feynman rules of SCET follow from the Lagrangian (\ref{LSCET}) in the usual way.
 
At subleading order in the expansion in powers of $\lambda$ new interaction vertices arise. The terms of ${\cal O}(\lambda)$ and ${\cal O}(\lambda^2)$ have been constructed in \cite{Beneke:2002ph}. Of particular importance to our discussion below is the coupling of a collinear fermion to an ultra-soft fermion $q_{us}$, which enters at first order in $\lambda$. The relevant effective Lagrangian reads
\begin{equation}\label{Lxiq} 
   {\cal L}_{\xi q,n}^{(1)}
   = \bar\xi_n(x)\,i\Dsl_{\perp c}\,W_n(x)\,q_{us}(x_-) + \mbox{h.c.} \,,
\end{equation}
where $W_n$ is the collinear Wilson line introduced in (\ref{Wn}), and the ultra-soft quark field has power counting $q_{us}\sim\lambda^3$. The Lagrangians (\ref{LSCET}) and (\ref{Lxiq}) can be written for any collinear sector of the theory.

\subsubsection*{Endpoint singularities in collinear contributions}

\begin{figure}
\begin{center}
\includegraphics[width=0.28\textwidth]{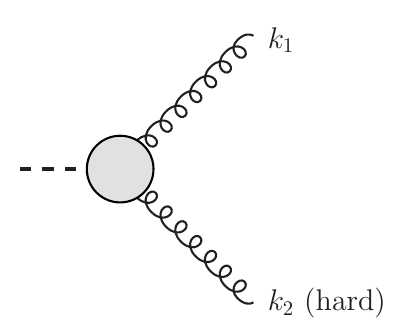} \quad
\includegraphics[width=0.25\textwidth]{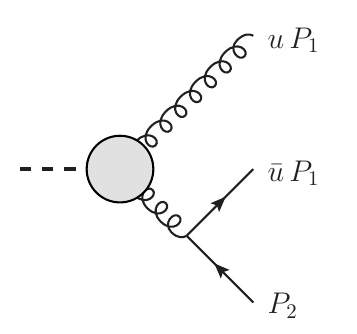} 
\end{center}
\vspace{-2mm}
\caption{\label{fig:offshellvertex} 
Vertex function connecting $S$ with a collinear gluon and a hard gluon (left), and the corresponding hard matching contribution to the Wilson coefficients in (\ref{poleterms}) (right).}
\end{figure}

For new-physics models in which the Wilson coefficients of the leading SCET$_{\rm BSM}$ operators in (\ref{Leff2}) are non-zero, one can show on general grounds that the Wilson coefficients $\bm{C}_{F_L\bar f_R\,\phi}^{(i)}(u,\mu)$, $\bm{C}_{F_L\bar F_L A}(u,\mu)$, and $\bm{C}_{f_R\bar f_R A}(u,\mu)$ are singular in the limit $u\to 1$. The origin of these singularities can be understood as follows. When integrating out some heavy degrees of freedom generates the operators in (\ref{lam2ops}), the same UV physics will also generate corresponding vertices in which one of the two outgoing collinear lines is replaced by a line carrying a hard momentum. Consider, for example, the vertex shown on the left-hand side in Figure~\ref{fig:offshellvertex} (a corresponding graph exists with $n_1$ and $n_2$ interchanged). If we denote the momentum of the $n_1$-collinear gluon by $k_1=u P_1$, then the hard gluon carries momentum $k_2=P_2+(1-u)P_1$. The vertex function can then be written in the form
\begin{equation}
   M \left[ C_{GG}(u,M_S,M,\mu)\,g_{\alpha\beta}^\perp
    + \widetilde C_{GG}(u,M_S,M,\mu)\,\epsilon_{\alpha\beta}^\perp \right] g_s^2\,\delta_{ab} \,,
\end{equation}
where the dependence on $u$ enters through the invariants $2k_1\cdot k_2=u M_S^2$ and $k_2^2=(1-u) M_S^2$. Clearly, for $u\to 1$ we recover
\begin{equation}
   \lim_{u\to 1}\,C_{GG}(u,M_S,M,\mu) = C_{GG}(M_S,M,\mu) \,,
\end{equation}
and likewise for $\widetilde C_{GG}$, where $C_{GG}$ and $\widetilde C_{GG}$ are the coefficients in the effective Lagrangian (\ref{Leff2}).\footnote{Beyond tree level this relation is more complicated. The coefficient on the left-hand side can contain hard loop corrections $\sim(\mu^2/k_2^2)^{n\epsilon}$, which are absent in the coefficient on the right-hand side.} 
Consider now the diagram shown on the right-hand side in Figure~\ref{fig:offshellvertex}, which yields the following hard matching contributions to the Wilson coefficients (omitting some arguments): 
\begin{equation}\label{poleterms}
\begin{aligned}
   \Delta\bm{C}_{Q_L\bar Q_L G}(u,\mu) 
   &= \frac{M^2}{M_S^2}\,\frac{g_s^2(\mu)}{1-u}
    \left[ C_{GG}(u,\mu) - i\widetilde C_{GG}(u,\mu) \right] , \\
   \Delta\bm{C}_{q_R\bar q_R G}(u,\mu) 
   &= \frac{M^2}{M_S^2}\,\frac{g_s^2(\mu)}{1-u}
    \left[ C_{GG}(u,\mu) + i\widetilde C_{GG}(u,\mu) \right] .
\end{aligned}
\end{equation}
This produces poles at $u=1$, whose residues are given in terms of the coefficients $C_{GG}$ and $\widetilde C_{GG}$ in the effective Lagrangian (\ref{Leff2}). At first sight, these give rise to endpoint-divergent integrals $\int_0^1\!dw\,\frac{1}{1-w}$ when inserted into (\ref{eq78}). 

To see how these integrals are cured, we need to look at the relevant operator mixing contribution in more detail. Consider the one-loop contributions to the $S\to q_L^i(k_1)\,\bar q_R^j(k_2)\,\phi^*(0)$ decay amplitude, where the scalar field carries zero momentum. We include multiplicative radiative corrections to the matrix element of the operator $O_{Q_L\bar q_R}^{\,ij}$ as well as the mixing contribution shown by the first two diagrams in Figure~\ref{fig:opmix}. Before renormalization, i.e.\ written in terms of bare Wilson coefficients, we find
\begin{equation}\label{step1}
\begin{aligned}
   {\cal M}(S\to q_L^i\bar q_R^j\phi^*)
   &= \frac{1}{M}\,\bigg[ Z_{Q_L\bar q_R}^{-1}\,C_{Q_L\bar q_R}^{ij} 
    - \int_0^1\!dw\,N_\epsilon(w) \left( \frac{\mu^2}{-k_1^2} \right)^\epsilon 
    Y_q^{ik}\,C_{q_R\bar q_R G}^{kj}(w) \\
   &\hspace{13mm}\mbox{}- \int_0^1\!dw\,N_\epsilon(w) \left( \frac{\mu^2}{-k_2^2} \right)^\epsilon 
    \big( C_{Q_L\bar Q_L G}^\dagger(w) \big)^{ik}\,Y_q^{kj} \bigg]
    \left\langle q_L^i\bar q_R^j\phi^*|O_{Q_L\bar q_R}^{ij}|S\right\rangle_{\rm tree} ,
\end{aligned}
\end{equation}
where (here and below we omit the ``$-i0$'' regulator in the arguments of the logarithms)
\begin{equation}
\begin{aligned}
   Z_{Q_L\bar q_R}^{-1} 
   &= 1 + \frac{C_F\alpha_s}{\pi} \left[ \frac{1}{2\epsilon^2} + \frac{1}{2\epsilon} 
    \left( \ln\frac{\mu^2}{-M_S^2} + \frac32 \right) \right] , \\
   N_\epsilon(w) & = e^{\epsilon\gamma_E}\,\frac{C_F\alpha_s}{2\pi}\,(1-\epsilon)\,\Gamma(\epsilon)\, 
    w\hspace{0.3mm} \big( w(1-w) \big)^{-\epsilon} \,.
\end{aligned}
\end{equation}
Naively expanding $N_\epsilon(w)$ as $N_\epsilon(w)=C_F\alpha_s\hspace{0.3mm} w/(2\pi\epsilon)+\mbox{``finite terms''}$ reproduces the mixing terms shown in (\ref{eq78}). However, in the presence of the poles at $u=1$ in (\ref{poleterms}), such an expansion does not capture all the $1/\epsilon$ singularities. Let us split up the Wilson coefficients in two terms, such that
\begin{equation}\label{Csubtract}
   C_{q_R\bar q_R G}^{ij}(w) 
   = \frac{M^2}{M_S^2}\,\frac{g_s^2}{1-w} \left( C_{GG} + i\widetilde C_{GG} \right) 
    \delta^{ij} + \bar C_{q_R\bar q_R G}^{ij}(w) \,,
\end{equation}
and similarly for $C_{Q_L\bar Q_L G}^\dagger(w)$. The subtracted coefficients $\bar C_{q_R\bar q_R G}^{ij}(w)$ and $\bar C_{Q_L\bar Q_L G}^{ij}(w)$ are integrable at $w=1$. We then obtain from (\ref{step1}) 
\begin{equation}\label{step2}
\begin{aligned}
   {\cal M}(S\to q_L^i\bar q_R^j\phi^*)
   &= \frac{1}{M}\,\Bigg[ Z_{Q_L\bar q_R}^{-1}\,C_{Q_L\bar q_R}^{ij} 
    - \frac{C_F\alpha_s}{2\pi\epsilon} \int_0^1\!dw\,w \left[ Y_q^{ik}\,\bar C_{q_R\bar q_R G}^{kj}(w) 
    + \big( \bar C_{Q_L\bar Q_L G}^\dagger(w) \big)^{ik}\,Y_q^{kj} \right] \\
   &\hspace{12mm}\mbox{}- \frac{C_F\alpha_s}{2\pi}\,e^{\epsilon\gamma_E}\,
    (1-\epsilon)\,\Gamma(\epsilon)\,\frac{\Gamma(2-\epsilon)\,\Gamma(-\epsilon)}{\Gamma(2-2\epsilon)}
    \left[ \left( \frac{\mu^2}{-k_1^2} \right)^\epsilon 
    + \left( \frac{\mu^2}{-k_2^2} \right)^\epsilon \,\right] \\
   &\hspace{16mm}\mbox{}\times
    \frac{g_s^2 M^2}{M_S^2} \left( C_{GG} + i\widetilde C_{GG} \right) Y_q^{ij} \Bigg]
    \left\langle q_L^i\bar q_R^j\phi^*|O_{Q_L\bar q_R}^{ij}|S\right\rangle_{\rm tree} .
\end{aligned}
\end{equation}
It follows that, in the $\overline{\rm MS}$ subtraction scheme, the bare Wilson coefficient $C_{Q_L\bar q_R}^{ij}$ receives the counterterms
\begin{equation}\label{step3}
\begin{aligned}
   C_{Q_L\bar q_R}^{ij} \big|_{\rm ren}
   &= Z_{Q_L\bar q_R}^{-1}\,C_{Q_L\bar q_R}^{ij} 
    - \frac{C_F\alpha_s}{2\pi\epsilon} \int_0^1\!dw\,w \left[ Y_q^{ik}\,\bar C_{q_R\bar q_R G}^{kj}(w) 
    + \big( \bar C_{Q_L\bar Q_L G}^\dagger(w) \big)^{ik}\,Y_q^{kj} \right] \\
   &\hspace{12mm}\mbox{}+ \frac{C_F\alpha_s}{2\pi} \left[ \frac{2}{\epsilon^2}
    + \frac{1}{\epsilon} \left( \ln\frac{\mu^2}{-k_1^2} + \ln\frac{\mu^2}{-k_2^2} \right) \right] 
    \frac{g_s^2 M^2}{M_S^2} \left( C_{GG} + i\widetilde C_{GG} \right) Y_q^{ij} .
\end{aligned}
\end{equation}
The endpoint singularities are regularized in this expression and give rise to the double poles in $1/\epsilon$; however, the appearance of the collinear logarithms is worrisome, as it would indicate a sensitivity of the associated anomalous dimension to infrared scales. 

\subsubsection*{Contribution from the exchange of an ultra-soft quark}

This dependence is cancelled by the contribution from a loop diagram involving the exchange of an ultra-soft quark between the two collinear sectors, shown in the bottom row of Figure~\ref{fig:opmix}. In this graph the $Sgg$ vertex descents from the ${\cal O}(\lambda^2)$ effective Lagrangian (\ref{Leff2}). It is combined with two insertions of the subleading SCET Lagrangian (\ref{Lxiq}), which couples a collinear fermion to a collinear gauge field and an ultra-soft quark. More accurately, the diagram arises from the subleading-power operator
\begin{equation}\label{Tproduct}
   T\,\Big\{ O_{GG}(x), i\!\int\!d^4y\,{\cal L}_{\xi q,n_1}^{(1)}(y), 
    i\!\int\!d^4z\,{\cal L}_{\xi q,n_2}^{(1)}(z), i\!\int\!d^4w\,
    {\cal L}_{\bar q\Phi q}^{(-1)}(w) \Big\} \,, 
\end{equation}
and similarly with $\widetilde O_{GG}$ instead of $O_{GG}$. The Lagrangian
\begin{equation}\label{Lfunny}
   {\cal L}_{\bar q\Phi q}^{(-1)} 
   = - \left( \bar q_{us,L}\,\bm{Y}_q\,\Phi_0\,q_{us,R} + \mbox{h.c.} \right) 
\end{equation}
describes the coupling of ultra-soft quarks to the zero-momentum scalar field $\Phi_0$. With $q_{us}\sim\lambda^3$ and $\Phi_0\sim\lambda$, and taking into account that the ultra-soft measure scales as $d^4x_{us}\sim\lambda^{-8}$, it follows that this Lagrangian contributes terms of ${\cal O}(\lambda^{-1})$ to the action. This lifts the operator in (\ref{Tproduct}) from the naive expectation ${\cal O}(\lambda^4)$ to ${\cal O}(\lambda^3)$.\footnote{One might worry that multiple insertions of the Lagrangian (\ref{Lfunny}) can promote the operator to even lower order in $\lambda$. However, graphs with such multiple insertions do not produce UV poles and are scaleless when evaluated on shell. If we would introduce soft mass-mode fields instead of ultra-soft fields, then the coupling of the soft quark to the scalar doublet is a leading-power interaction, while the coupling of a soft quarks to a collinear quark and gluon in (\ref{Lxiq}) appears at ${\cal O}(\lambda^{1/2})$. Also in this case the operator (\ref{Tproduct}) is of ${\cal O}(\lambda^3)$.}

Evaluating the contribution of the operator (\ref{Tproduct}) to the matrix element in (\ref{step1}), we obtain an extra contribution inside the bracket on the right-hand side of (\ref{step2}), which reads
\begin{equation}\label{uscontr}
   - \frac{C_F\alpha_s}{2\pi}\,e^{\epsilon\gamma_E}\,(1-\epsilon)\,\Gamma(\epsilon)\,
   \frac{\pi}{\sin\pi\epsilon} \left( \frac{\mu^2(-M_S^2)}{(-k_1^2)(-k_2^2)} \right)^\epsilon 
   \frac{g_s^2 M^2}{M_S^2} \left( C_{GG} + i\widetilde C_{GG} \right) Y_q^{ij} \,.
\end{equation}
This term has the effect of removing the collinear logarithms in expression (\ref{step3}) and replacing them by a logarithm of the hard scale. We thus obtain the final result
\begin{equation}
\begin{aligned}
   C_{Q_L\bar q_R}^{ij} \big|_{\rm ren}
   &= Z_{Q_L\bar q_R}^{-1}\,C_{Q_L\bar q_R}^{ij} 
    - \frac{C_F\alpha_s}{2\pi\epsilon} \int_0^1\!dw \left[ Y_q^{ik}\,\bar C_{q_R\bar q_R G}^{kj}(w) 
    + \big( \bar C_{Q_L\bar Q_L G}^\dagger(w) \big)^{ik}\,Y_q^{kj} \right] \\
   &\hspace{12mm}\mbox{}+ \frac{C_F\alpha_s}{2\pi} \left[ \frac{1}{\epsilon^2}
    + \frac{1}{\epsilon} \left( \ln\frac{\mu^2}{-M_S^2} - 1 \right) \right] 
    \frac{g_s^2 M^2}{M_S^2} \left( C_{GG} + i\widetilde C_{GG} \right) Y_q^{ij}
\end{aligned}
\end{equation}
for the counterterms. From this expression, it is straightforward to derive the RG evolution equation (\ref{rgefinal}).

\end{appendix}

\end{document}